\documentclass[manuscript]{aastex}
\usepackage[usenames]{color}
\usepackage{graphicx}
\usepackage{epstopdf}
\usepackage{xcolor}
\usepackage{lineno}
\shorttitle{Ion Heating in Inhomogenous Expanding Solar Wind Plasma}
\shortauthors{Ozak, Ofman, Vi\~nas}

\begin{document}

%
%

\title{Ion Heating in Inhomogenous Expanding Solar Wind Plasma: The Role of Parallel and Oblique Ion-Cyclotron Waves}
\author{N. Ozak\altaffilmark{1}, L. Ofman\altaffilmark{2,3,4}, A.-F. Vi\~nas\altaffilmark{3}}
\altaffiltext{1}{Department of Earth and Planetary Sciences, Weizmann Institute of Science, Rehovot, Israel}
\altaffiltext{2}{Catholic University of America, Washington, DC 20064, USA} 
\altaffiltext{3}{NASA Goddard Space Flight Center, Greenbelt, MD 20771, USA} 
\altaffiltext{4}{Visiting, Department of
Geophysics and Planetary Sciences, Tel Aviv University, Tel Aviv 69978, Israel}

\begin{abstract}
Remote sensing observations of coronal holes show that heavy ions are hotter than protons and their temperature is anisotropic. In-situ observations of fast solar wind streams provide direct evidence for turbulent Alfv\'en wave spectrum, left-hand polarized ion-cyclotron waves, and He$^{++}$ -- proton drift in the solar wind plasma, which can produce temperature anisotropies by resonant absorption and perpendicular heating of the ions. Furthermore, the solar wind is expected to be inhomogeneous on decreasing scales approaching the Sun. We study the heating of solar wind ions in inhomogeneous plasma with a 2.5D hybrid code. \textbf{We include the expansion of the solar wind in an inhomogeneous plasma background, combined with the effects of a turbulent wave spectrum of Alfv\'enic fluctuations and initial ion-proton drifts}. We study the influence of these effects on the perpendicular ion heating and cooling and on the spectrum of the magnetic fluctuations in the inhomogeneous background wind. We find that inhomogeneities in the plasma lead to enhanced heating compared to the homogenous solar wind, and the generation of significant power of oblique waves in the solar wind plasma. The cooling effect due to the expansion is not significant \textbf{for super-Alfv\'enic drifts}, and is diminished further when we include an inhomogenous background density. We reproduce the ion temperature anisotropy seen in observations and previous models, which is present regardless of the perpendicular cooling due to solar wind expansion. We conclude that small scale inhomogeneities in the inner heliosphere can significantly affect resonant wave ion heating.

\end{abstract}

%
%

\section{Introduction}
The heating and the acceleration of the solar wind multi-ion plasma is still a poorly understood phenomenon, despite the fact that it has been studied extensively for decades by satellite observations and theoretical models. In situ and remote sensing observations at 0.29 AU and beyond by Helios and Ulysses spacecraft have found non-thermal features in the ion velocity distributions \citep[i.e.,][]{Mar82a,Fel96,Neu96}. For instance, proton distributions often appeared double-peaked,  He$^{++}$ ions drift at the local Alfv\'en speed relative to protons, and heavy ions usually appear hotter and flow faster than protons in the fast solar wind streams. Stronger perpendicular heating in ions than in protons has been observed in fast wind streams \citep[e.g.,][]{Mar82a,Mar82b,Ger12}. From pure adiabatic expansion of the solar wind plasma one would expect that $T_{\perp} < T_{||}$ due to the conservation of magnetic moment of the expanding ions in a decreasing radial magnetic field. However, remote sensing close to the sun and in-situ observations at $<0.3$ AU of fast solar wind \textbf{reveal that for ions $T_\perp > T_{||}$}. This anisotropy has been attributed as indirect evidence for the presence of ion-cyclotron waves, as it has been suggested that resonant absorption of ion cyclotron waves heats and accelerates the ions in the solar wind \citep[e.g.,][]{AM92,TM97,Li99,OVG01,HI02,Jia14,Omi14}. Furthermore, observations by Helios, ACE, Wind, and Ulysses show that magnetic fluctuations in fast solar wind streams can be fitted by simple power laws, providing clues on the possible heating mechanism \citep[e.g.,][]{Bav82,Gol95,Pod06,Vas07,Sal09}. 

Measurements of the magnetic spectrum fluctuations provide further clues to the turbulence and dissipation processes involved in the solar wind plasma heating. For instance, observations of Alfv\'enic fluctuations in the solar wind by Helios, Ulysses, ACE, and Wind spacecraft show that the fluctuations follow power laws of $f^{-1}$ (where $f$ is the frequency in spacecraft frame) at very low frequencies, $f^{-5/3}$ suggestive of Kolmogorov turbulence in the inertia range, and steeper slopes in the dissipation range near proton gyroresonant frequency. The Solar Probe Plus mission, currently under development by NASA, is scheduled to launch in 2018, and will measure the solar wind properties as close as 9.5 R$_S$ from the Sun (where $R_s$ is the solar radius $\sim 7 \times 10^8$ m). In this region the solar wind plasma is still accelerating and it is expected to be inhomogeneous due to the cross-field density and velocity structures of the corona. The effect of plasma inhomogeneity is considered in the present study. 

Previous work by \citet{OV07} on homogeneous plasma heating with a 2D hybrid model shows that non linear effects are important in the heating of the homogenous plasma early in the simulation. They find that a driven Alfv\'en broadband spectrum causes perpendicular heating of heavy ions. However, they conclude that oblique waves do not play an important role in the heating of homogenous plasma. Recently, \citet{OVM14} studied the heating of expanding homogenous solar wind plasma with a two-dimensional hybrid model. They find that the solar wind expansion has little effect on the preferential ion heating but that it leads to faster evolution of magnetosonic drift instability. Several  2D hybrid models of homogeneous solar wind plasma heating by a spectrum of ion cyclotron waves \textbf{include studies} by \citet{Gar01,Gar03,Gar06,Hel03,Hel05,Omi14,MOV14}. Previous works that study the effect of the expansion on homogeneous solar wind plasma heated by ion-cyclotron waves with 1D and 2D hybrid models include \citet{LVG01,OVM11,MVO13,Hel13}, and \citet{OVM14}.

Work by \citet{Ofm10a} and \citet{OVM11} investigate with a 2D hybrid model the effects of a turbulent spectrum on the heating of solar wind plasma in an inhomogenous density background. They find that He$^{++}$ ions are heated more efficiently by the Alfv\'enic wave spectrum below the proton gyroperiod compared to the homogeneous background solar wind. In this work we extend their study to an expanding-box 2.5D hybrid model. We explore the effect of sub- and super-Alfv\'enic ion relative drifts as well as a turbulent source spectrum on the perpendicular heavy-ion heating in an inhomogeneous plasma with various degrees of cross-field density gradient. We investigate the collisionless heating processes in the expanding solar wind using plasma parameters appropriate close to the Sun ($\sim10R_s$). We find that the inhomogeneity in the plasma generates oblique waves due to refraction of an initially parallel wave spectrum, which contribute to the heating of the solar wind plasma, while left-hand polarized parallel propagating waves still play the main role in the perpendicular ion heating.  

The paper is organized as follows: in Section~\ref{model:sec} we present the details of our numerical hybrid model, in Section~\ref{num_res:sec} we show the numerical results, and Section~\ref{conc:sec} is devoted to discussion and conclusions.

\section{Model}
\label{model:sec}
For our calculations we use a 2.5D hybrid model initially developed by \citet{WO93}, adapted for 2D modeling of waves and beam heating of solar wind plasma by \citet{OV07} and later parallelized by \citet{Ofm10a}.  The hybrid model allows the study of ion dynamics by treating the ions kinetically using the Particle-In-Cell (PIC) technique in a two-dimensional spatial domain (allowing parallel and obliquely propagating waves), while considering the electrons as a charge-neutralizing background fluid.  In addition, the model follows the three components of the particle velocity and of the electric and magnetic fields. We use a Cartesian coordinate system with the $x$-direction along the uniform background magnetic field $B_0\hat{x}$ and assume periodic boundary conditions in $x$ and $y$. The simulations follow more than $8\times 10^6$ particles, which are initialized as 127 particles/cell/species in a 2D spatial grid of 128 by 256 computational cells. This number of particles ensures that the velocity phase space is well sampled, and the calculated velocity distribution functions (VDF) are not significantly affected by statistical noise. The size of the modeled grid has units of 1.5$\Delta$ in the $x$-direction, and 0.75$\Delta$ in the $y$-direction, where $\Delta = c/\omega_{pp}$ is the proton inertia length and $\omega_{pp}$ is the plasma frequency of the protons. The inhomogeneity across the field is well resolved, and the wavenumber range in the model covers the low frequency (non-resonant) part of the dispersion relation as well as waves in the dissipation range \citep[e.g.,][]{OV07}. 

\begin{figure}
\begin{tabular}{c}
\includegraphics[scale=0.75]{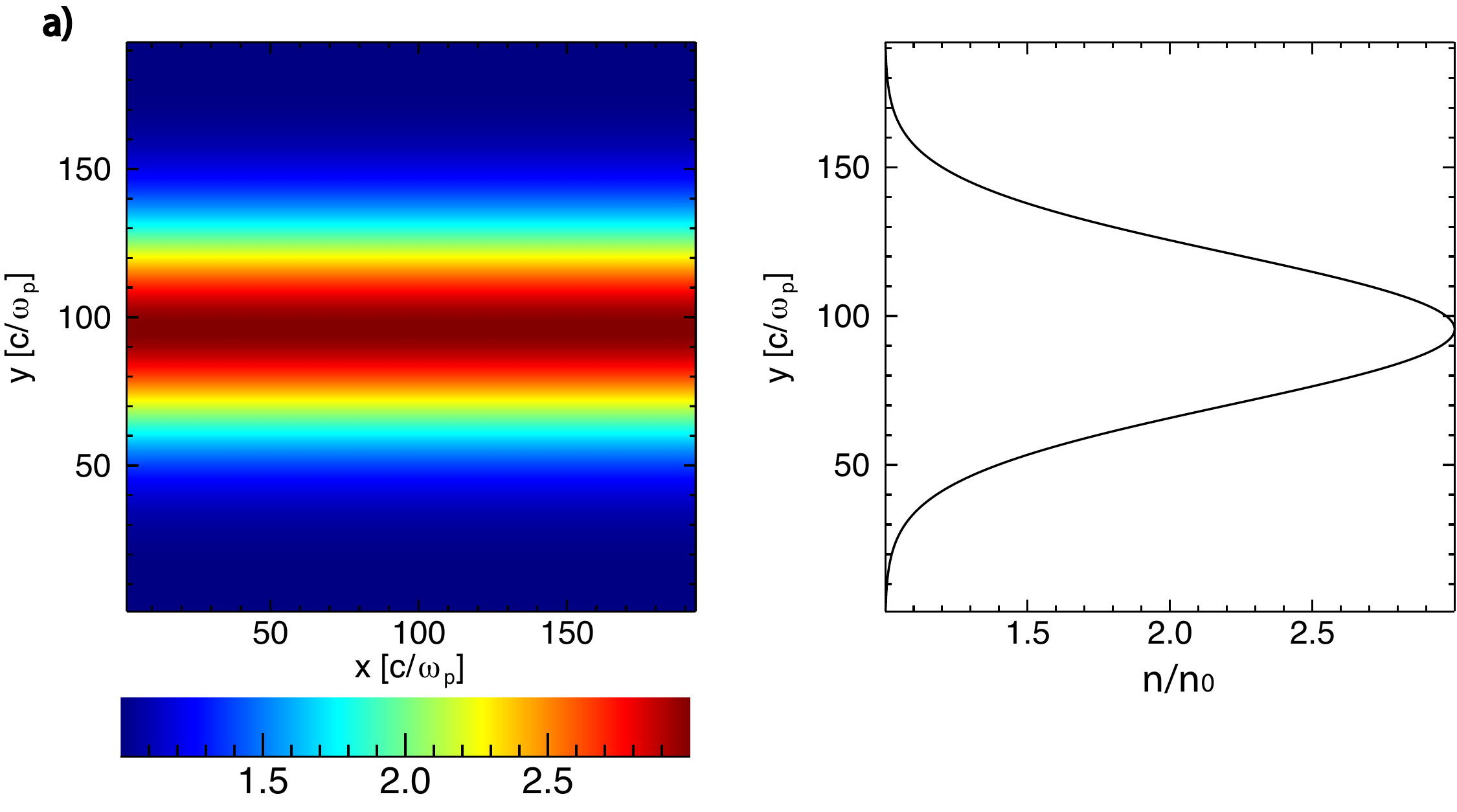} \\
\includegraphics[scale=0.75]{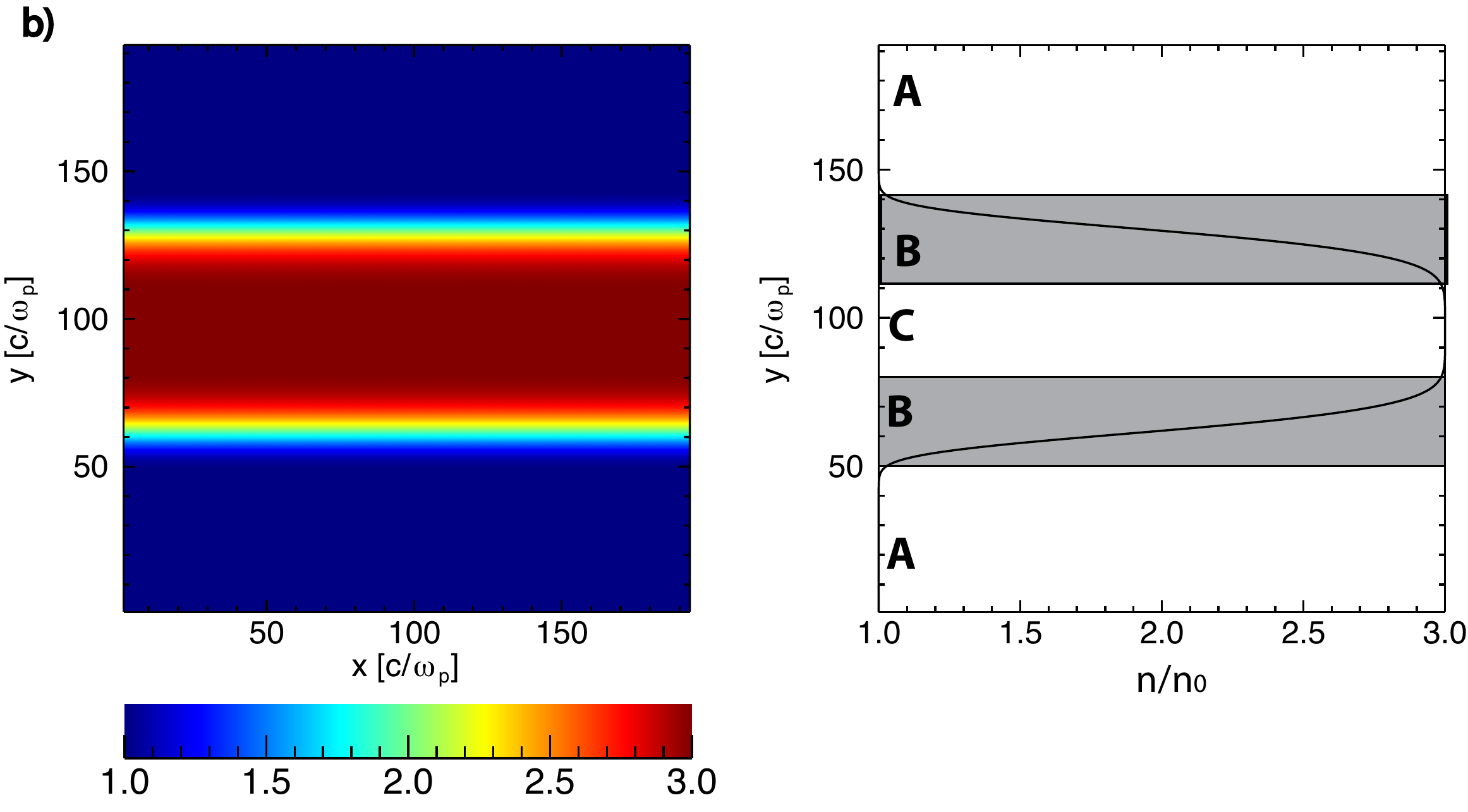} \\
 \end{tabular}
 \caption{Initial density distributions of protons and He$^{++}$ ions with Gaussian inhomogeneity (top, $q=2$) and sharp inhomogeneity (bottom, $q=6$).}\label{initial_density}
 \end{figure}

In this work we study the effect of an inhomogenous proton and He$^{++}$ background density on the dissipation of Alfv\'enic fluctuations in the solar wind plasma (the electron density is determined by the quasi-neutrality condition $n_e=2n_i+n_p$. We initialize our model calculations with a normalized proton and ion background density given by:
\begin{equation}
n_{k}(y) = n_{0,k}\left[1+n_{0m}e^{-(\frac{y-y_0}{w})^q}\right],
\end{equation}
where $k$ indicates the particle species (protons or He$^{++}$ ions) and $y_0$ is the location of the central symmetry axis. $n_{0m}$ is the amplitude of the inhomogeneity, which is taken to be $n_{0m}=2$ in this study (\citet{Ofm10a} performed a parametric study of $n_{0m}$ in the range 0 -- 4 and of other parameters in non-expanding solar wind plasma, where $n_{0m}=0$ represents homogeneous background). The parameter $q$ is the power that determines the sharpness of the transition between low and high density regions, and $w=38.4\Delta$ gives the width of the high-density region. Note that the inhomogeneity does not introduce a net current in this model due to the massless treatment of the electrons. Here we compare the effects of a sharp ($q=6$) and a less sharp Gaussian inhomogeneity ($q=2$) on the plasma heating. 
In the hybrid code each modeled particle is a ``superparticle'' that corresponds to large number of physical particles located in the same phase--space position. The density normalization determines the exact number of real particles that are represented by a superparticle in the code. The normalization factors in the model are $n_{0,p}=0.9n_e$ and $n_{0,He^{++}}=0.05n_e$, characteristic of the solar wind composition, and the initial system has no net currents. Figure \ref{initial_density} shows the initial spatial distributions of protons and He$^{++}$ ions used in our model for the Gaussian ($q=2$) and sharper than Gaussian ($q=6$) inhomogeneities. Both ion plasma species have the same initial spatial distributions. We divide the sharp inhomogeneity case into three distinct regions: an external homogeneous region (region A in Figure \ref{initial_density}b), an inhomogenous region (shaded region B in Figure \ref{initial_density}b), and an internal homogenous region (region C in Figure \ref{initial_density}b).

In anticipation of the Solar Probe Plus mission observations close to the Sun we choose our modeling parameters to represent the solar wind conditions at $\sim10R_s$. The initial low-$\beta$ regime parameters are $\beta_e = 0.021$ and $\beta_{p||}=\beta_{He^{++}{||}}=0.041$. The particles in the model have isotropic Maxwellian velocity distributions initially. We obtain the solution to the equations of motion by advancing the particles and the fields in time using the Rational-Runge-Kutta (RRK) method \citep{Wam78} and the pseudo-spectral Fourier method for spatial solution of the fields on the spatial grid \citep[e.g.,][]{Ter86}.

\subsection{Modeling Solar Wind Expansion}
As part of this study we investigate the effect of the expansion of the solar wind plasma on the temperature anisotropies and velocity distribution functions. The expansion model used was initially proposed for MHD equations by \citet{GV96} and was later implemented to hybrid 1D and 2D models \citep{LVG01,Hel03}. Our version of the model has been implemented in previous studies of solar wind expansion in 1D hybrid modeling with two ion species \citep{OVM11,MVO13} and in a 2D hybrid model of homogenous H$^{+}$ -- He$^{++}$ solar wind plasma \citep{OVM14,MOV14}. For this study we employ the same 2D expanding box model used previously, but include instead an inhomogeneous background plasma with density gradient across the background magnetic field direction. In brief, the expanding model can be summarized as follows: 
\begin{enumerate}
\item the model assumes radial expansion of the solar wind at a distance $R_0$ from the Sun with a constant bulk velocity $U_0\hat{e_r}$. 
\item The equations of motion and the fields are modified by a dimensionless expansion factor $a(t)=1+\epsilon t$, where $\epsilon=U_0/R_0$ is the expansion parameter and the $t$ is in units of $\Omega_p^{-1}$, such that $\epsilon \ll 1$ and higher order terms of $\epsilon$ are neglected. In our model we choose $\epsilon = 10^{-4}, 10^{-3}$, due to computational limitations, compared to a real solar value $\epsilon\sim10^{-5}$ near 10$R_S$. The background field and the corresponding ion-cyclotron frequency decrease as 1/$a^2$.
\item The coordinate system is transformed, such that the $x$-direction undergoes Galilean transformation and the $y$- and $z$- coordinates undergo stretching with time.
\end{enumerate}
We refer the reader to \citet{LVG01}, \textbf{\citet{Hel03}, \citet{Hel05},}  \citet{OVM11,OVM14} and \citet{MVO13,MOV14} for further details on the expanding box model and its implementation. 

\section{Numerical Results}
\label{num_res:sec}
We divide our study in two parts: in the first part, we model the effect of the expansion in the inhomogeneous solar wind plasma and include an initial beam of proton-He$^{++}$ drift with a super- or sub-Alfv\'enic relative drift velocity. In the second part, we include a turbulent spectrum of driven circularly polarized Alfv\'{e}nic/cyclotron waves in addition to the initially drifting ion beam. In each of these setups the super-Alfv\'enic drift and/or the spectrum provide sources of free energy \textbf{leading to ion-cyclotron waves} and self-consistent secondary wave spectrum. We calculate the waves' impact on the ion heating and on the temperature anisotropy in the model. We list the initial parameters of the simulations for all cases chosen in this study in Table \ref{par_table}.

\subsection{Effects of Solar Wind Expansion in Plasma with Initial Ion Drift}
In this section we apply a beam of He$^{++}$ ions with an initial drift with respect to the protons. First, we analyze the effect of a sub-Alfv\'enic ion drift $V_d = 0.5 V_A$, where $V_A= (4\pi\rho)^{1/2}$ is the Alfv\'en speed. We employ different expansion factors: $\epsilon=0$ (no expansion), $\epsilon = 10^{-4}$ and $10^{-3}$ demonstrating the effects of slow and moderate expansions (Cases 1 and 2). Figure~\ref{subalfv_gauss} shows the calculated temporal evolution of the temperature anisotropies for the Gaussian and sharp inhomogeneity cases respectively. Without the expansion there is a slight increase ($\sim 5\%$) in the ion anisotropy and a comparable decrease in the proton anisotropy by the end of the run. \textbf{This increase in the ion anisotropy is likely due to numerical heating, which is controlled in the code by looking at particle energy conservation. We refer the reader to \cite{BL91} and \cite{W03} for further reading on particle energy testing in PIC and hybrid codes, respectively. Our diagnostic confirmed that the particle energy is conserved reasonably well (typically within 5\%) throughout the run.} When the expansion factor is small ($\epsilon=10^{-4}$) the ion temperature anisotropy remains constant and the proton temperature anisotropy decreases about $10\%$. However, once moderate expansion is included ($\epsilon = 10^{-3}$) the temperature anisotropy of protons and He$^{++}$ ions decreases quickly with time by 35\% and 40\% at the end of the run, respectively. This is due to the stretching of the perpendicular coordinates and the corresponding decrease of the ions' perpendicular velocities that translates to cooler perpendicular VDFs of the ions. The ion drift in the parallel direction remains constant throughout the simulation and the sharpness of the background inhomogeneity does not have a significant effect on the evolution of the temperature anisotropy in this case. 

For comparison, we repeat the calculations using an initially super-Alf\'enic ion drift with $V_d= 2V_A$ (Cases 3 and 4). Super-Alf\'enic  drifts are unstable to ion cyclotron instability leading to perpendicular ion heating \citep{DGW99,Ara02,XOV04,OV07}. Figures \ref{superalfv_gauss} and \ref{superalfv_sharp} show our results for the Gaussian and sharp inhomogeneities. \textbf{The initial drift drives a magnetosonic instability that leads to a fast increase in the He$^{++}$ ion temperature anisotropy reaching a peak of $\sim4$ in about $t = 40 ~\Omega ^{-1}$ for all expanding cases and both inhomogeneity sharpness cases. Once the peak is reached, the anisotropy decreases quickly as the He$^{++}$ emit ion-cyclotron waves due to their own ion-cyclotron instability, as well as due to some gradual parallel heating (see Figure \ref{energies_exp_comp}). Consequently, the ion drift relaxes due to the energy lost by the emission of the waves and the protons are heated rapidly in the perpendicular direction due to resonance.} The proton temperature anisotropy grows reaching 1.5 in the Gaussian inhomogeneity case and $\sim1.6$ in the sharp inhomogeneity case within $t = 50~ \Omega^{-1}$. Parallel heating also takes place by velocity space diffusion, as evident in Figure~\ref{energies_exp_comp}. 
When the expansion is taken into account, initially there is a drop in the proton temperature anisotropy due to the perpendicular cooling effect from the expansion, which then increases to slightly lower maximum values of the temperature anisotropies at a later time, compared to the non-expanding case. At the end of the run the proton temperature anisotropy for the Gaussian case has not reached equilibrium yet and continues to increase. Similar increase in the anisotropy is also seen for cases with a homogeneous plasma density background \citep[e.g.,][]{OVM14}, or when the inhomogeneity gradient across the field is moderate like in the Gaussian inhomogeneity case. 

For the sharp inhomogeneity case ($q=6$) the temperature anisotropy reaches an equilibrium faster compared to the $q=2$ case as long as the expansion is kept at a slow rate. For the moderate expansion case shown ($\epsilon = 10^{-3}$) the proton temperature anisotropy is still increasing at the end of the simulation. The ion drift is also affected by the expansion, which causes a faster decrease to lower velocity values. Figures \ref{superalfv_gauss} and \ref{superalfv_sharp} also show a small peak in the ion drift, which correlates with a peak in the ion temperature anisotropy. This effect is caused by an increase in the parallel velocity of the protons, when the peak perpendicular heating is reached. The heating of the protons in the parallel direction then causes a $\sim$ 5 -- 7\% increase in the drift, which in turn heats the ions and increases the anisotropy momentarily by 20\%. The ion temperature anisotropy reaches an asymptotic value slightly larger than 1, while the protons continue to heat in the perpendicular direction, unless inhibited by a large expansion coefficient. It is evident from Figure~\ref{energies_exp_comp} that the slow solar wind expansion rate has little or no effect on the particle energies, while a moderate expansion rate has an increasing effect in the perpendicular component of the ions and a slight decreasing effect on the perpendicular energy of the protons. These results are in agreement with the previous studies of magnetosonic drift instability in inhomogeneous background solar wind plasma \citep[e.g.,][]{Ofm10a,OVM11,OVM14}. The expansion appears to mostly affect the proton temperature anisotropy (since the protons are not heated significantly by the instability) and speeds up the decrease of the relative He$^{++}$-proton drift. 

Figures \ref{vspaceGauss}a, \ref{vspaceGauss}b, and \ref{yz-phasespace} show the $V_y$ -- $V_z$ plane of the velocity phase-space for protons and He$^{++}$ ions for the super-Alfv\'enic drift case with gradual expansion ($\epsilon=10^{-4}$). Figure~\ref{vspaceGauss} conveys the results for the Gaussian background inhomogeneity (case 3b). Below each phase-space plot we show the velocity distribution in $V_z$ at $V_y=0$ with the corresponding best-fit Maxwellian distribution over-plotted with the dotted curve. The results show the proton distribution close to equilibrium at the end of the simulation (see Figure \ref{vspaceGauss}a). On the other hand, the He$^{++}$ ions perpendicular velocity distribution deviates from a Maxwellian in the core of the distribution. \textbf{This deviation effect is due to the drift instability, which affects the ions, and has been also observed in models without the inhomogeneity \citep{OVM14}. Although the distribution is isotropic in the perpendicular velocity phase space, at the tails it decreases faster than the best-fit Maxwellian distribution}. Figure~\ref{yz-phasespace} shows the results for a sharp background inhomogeneity (Case 4b). In the sharp inhomogeneity case we calculate and compare the velocity distribution in the three distinct regions as explained above: an inhomogenous region (shaded region B in Figure \ref{initial_density}b), an internal homogenous region (region C in Figure \ref{initial_density}b), and an external homogeneous region (region A in Figure \ref{initial_density}b). The proton distribution for all three regions (not shown) follows a Maxwellian distribution closely, as was shown previously in the Gaussian inhomogeneity case. The He$^{++}$ ion distributions in all three regions are circularly symmetric in the perpendicular plane. However, the distribution \textbf{only follows a Maxwellian at the core and the tails of the distribution decrease at different rates in each of the regions}. This effect may be attributed to the background inhomogeneity, as it is not seen in the results shown by \citet{OVM14}, which analyzes a homogeneous solar wind plasma case with similar parameters. At the tails the distribution falls faster than the Maxwellian in all there regions, similarly to the Gaussian inhomogeneity case. Figures \ref{vspaceGauss}c, \ref{vspaceGauss}d, and \ref{xz-phasespace} show the  $V_x$ --$V_z$ plane of the phase-space velocity distribution for both background inhomogeneity cases. Figure~\ref{vspaceGauss}c demonstrates the preferential perpendicular heating of the protons, although the difference with the best-fit Maxwellian distribution is not pronounced. From Figure~\ref{vspaceGauss}d and Figure~\ref{xz-phasespace} it is clear that the velocity distributions for the He$^{++}$ ions is distinctly non-Maxwellian regardless of the shape of the inhomogeneity. Moreover, the sharper inhomogeneity leads to a narrower beam-like VDFs for the ions. In fact, even the homogenous regions of the sharp inhomogeneity case exhibit a clear preferential heating. The external homogeneous region, region A, is a crescent shaped distribution with distinct wings. The inhomogeneous region, region B,  and the internal homogeneous region, region C, show a Maxwellian halo with a crescent shaped core. 

\subsection{Effects of Solar Wind Expansion in Plasma with Initial Ion Drift and Turbulent Spectrum}
For the second part of this study we use a temporally driven wave spectrum at the simulation boundary combined with a slow solar wind expansion ($\epsilon=10^{-4}$) to study the heating of the inhomogeneous plasma (Case 5). We keep our initial conditions the same as in the previous section with the initial ion drift, but with the addition of turbulent wave spectrum of circularly polarized Alfv\'en/cyclotron waves injected at the boundary (see \citet{Ofm10a}, Figure 4). The turbulent wave spectrum is imposed  by driving non-uniform magnetic fluctuations at the boundary given by \citep[see,][]{OVM14}:
\begin{equation} 
 	B_{z}(t,x=0,y) = B_{z0}\sum\limits_{i=1}^{N}a_i\sin(\omega_{i}t+ \Gamma_{i}(y)),
 \end{equation}
where $a_{i} = i^{-p/2}$ is the $i$'th mode amplitude, \textbf{$p$ is the given parameter that determines the slope of the power spectrum ($p=1$ in our study)}. N = 300, $\Delta\omega = (\omega_{N} - \omega_{i})/(N -1)$ is the frequency range. In the present study $\omega \in[0.06, 0.4]$ or $\omega \in[0.06, 0.9]$ (see Table 1). $\Gamma_{i}(y)$ is the value of the $y$-dependent random phase in the range $(0,2\pi)$ varied at each $y$ grid location, and $B_{z0}$ is the amplitude of the magnetic field fluctuation. This leads to the formation of circularly polarized wave spectrum  due to coupling with the $B_y$ component through the field equations, and the propagation of these waves into the interior of the computational domain.

For illustration purposes, we first calculate the temperature anisotropy evolution for cases with initial sub-Alfv\'enic drifts and Gaussian inhomogeneity. We show these results in Figure \ref{subalfv_turb}. On one hand, there is a large increase in the He$^{++}$ ion temperature anisotropy, when the turbulent spectrum is included. The eventual decrease of the temperature anisotropy is due to the additional parallel heating caused by phase-space diffusion of the energy supplied by the turbulent spectrum. On the other hand, the proton temperature anisotropy slightly decreases, and there seems to be little proton heating due to the turbulent spectrum. Nevertheless, the rate of decrease in the anisotropy is slower when turbulence is included. In comparison, Figure \ref{superalfv_turb} shows the results for an initial super-Alfv\'enic drift ($V_d= 2V_A$), which is unstable to the drift magnetosonic instability, for the Gaussian inhomogeneity. We find that the protons are affected more by the turbulence compared to He$^{++}$. For instance, at $t = 500~\Omega^{-1}_{p}$ the proton  temperature anisotropy for the Gaussian inhomogeneity case without turbulent spectrum is 1.6, while it is only 1.4 for the case that includes the input wave spectrum; a 13\% decrease. For the sharp inhomogeneity (not shown) case the difference is even larger: 1.7 compared to 1.3 for the turbulent spectrum case, a 24\% decrease. To better understand the cause of decrease in the proton temperature anisotropy we show the perpendicular and parallel \textbf{kinetic} energies for the modeled solar wind plasma (Figure \ref{energies}). Overall, we see that the inclusion of turbulent wave spectrum in the calculation causes an increase in the parallel and perpendicular energies for both ions and protons. In particular, the relative increase in the parallel energy of the protons due to the turbulence is the greatest ($\sim$ 30\%), which explains the decrease in the anisotropy seen in Figure \ref{superalfv_turb}. It is also worth noticing that when the plasma is homogeneous (case analyzed by \citet{OVM14}), there is no increase in the parallel energy of the protons due to the drift. Nonetheless, our inhomogeneous case shows a 25\% increase ($q=2$ case), and even a greater increase when the turbulent wave spectrum is included. This suggests an increased heating due to oblique waves produced by the inhomogeneity. For comparison, we calculated another example in Case 6 with a different frequency range: $\omega \in[0.06, 0.9]$ (not shown). We find that with the turbulent spectrum used in Case 6 the peak of the ion anisotropy is slightly lower and the proton anisotropy  is slightly larger but showing similar temporal evolution to Case~5. The decay of the drift was unaffected by the turbulent spectrum used.

Figure~\ref{powerspec_turb} shows the power spectrum of magnetic fluctuations evaluated in the middle of the high density region \textbf{at the end of the simulation ($\mathbf{t = 500 \Omega ^{-1}_{p}}$). The spectrum is produced by kinetic magnetosonic instability at the same time that left-hand polarized waves are resonantly absorbed and re-emitted by the He$^{++}$ and the protons. The plotted spectrum is the result of the balance between these processes and the non-linear wave-particle interactions}.  We compare the spectrum of Cases 3a, 3b and 5a, i.e., with initial drift only (red), with initial drift and expansion (blue), and initial drift, expansion and driven turbulent spectrum (black), respectively. The spectrum remains nearly flat until about $\omega=0.2~\Omega_p$ for all cases, \textbf{as this region includes the non-resonant range of the spectrum}. We show the best-fit for each spectrum in the declining region, chosen where the slope of the spectrum can be best-fitted with a power law. We find that for plasma with an initial drift \textbf{(but no initial turbulent spectrum)} there is no significant change in the slope of the power law, found to be $-2.2$. This value is close to that found by \citet{OVM14}: $-2.1$. The small difference between the slopes of the homogeneous and the inhomogeneous cases shows that the inhomogeneity has little effect on the power spectrum of magnetic fluctuations produced by the drift instability. However, when the initial turbulent spectrum was included as a boundary condition the slope of the power law for the magnetic fluctuations decreased to $-1.5$. This decrease can be explained by the less steep slope of the injected spectrum ($-1$) compared to the slope of the drift instability produced spectrum. Thus the power law of the combined magnetic fluctuations produced by the two processes is in between the two extreme values.

To further address and understand the effect on the magnetic fluctuations spectrum we show the two-dimensional spatial spectra for several cases considered. Figure \ref{2dpowerspec} shows the $k$-space spectra for the inhomogenous plasma with (1) an initial turbulent spectrum, (2) an initial super-Alf\'enic drift (Case 3a), (3) an initial super-Alf\'enic drift with expanding solar wind (Case 3b),  and (4) a case including all effects: turbulent wave spectrum, drift, and expansion (Case 5a). All results are shown at t = 250$\Omega^{-1}_p$, \textbf{i.e., not all cases have reached equilibrium}. From Figure \ref{2dpowerspec} we can see that the magnitude of the spectrum generated by the initial drifting plasma is greater than that generated by just the injected turbulent spectrum. Figures~\ref{2dpowerspec}a - \ref{2dpowerspec}c show one peak power concentrated in the parallel direction near $|k_y|=0$. However, Figure \ref{2dpowerspec}d shows two local peaks in the power spectrum, the largest one due to the drift and the lower peak due to the turbulent spectrum, identified by comparing to the previous cases. All cases show significant power in the $|{k_y}| \ge 0$ region of the $k_x-k_y$ plane, showing the presence of oblique waves in the spectrum. In comparison with results from \citet{OVM14}, we show that the inclusion of inhomogenous plasma in our model significantly enhances the production of the oblique waves in the plasma, likely due to refraction of the parallel propagating waves by the gradient of the background Alfv\'en wave phase speed. Figures \ref{2dpowerspec}c and \ref{2dpowerspec}d have not yet reached a fully relaxed state. Therefore, the peak is still shifted and has a negative $k_y$ value. As time evolves the drift instability is damped and only \textbf{one peak due to the injected spectrum remains visible at the end of the run of the case shown in Figure \ref{2dpowerspec}d.} 

\textbf{Figure \ref{dispersion} shows the dispersion relation for the waves obtained from the nonlinear 2.5D hybrid model in Case 3b ($V_d= 2V_A$, with solar wind expansion $\epsilon = 10^{-4}$, and Gaussian inhomogeneity). We obtain the dispersion relation by Fourier transforming in space and time the transverse magnetic fluctuations ($B_{\perp}$), which are obtained from the hybrid model simulations. Figure \ref{dispersion}a illustrates the dispersion for the parallel waves ($\omega$ vs. $k_x$). The non-resonant (right-hand polarized) and resonant (left-hand polarized) branches of the waves are clearly visible in the figure. These branches can be related to an analytical (Vlasov) dispersion relation (e.g. \cite{XOV04}, \cite{ODNV05}, \cite{OV07} and \cite{MOV14}). The non-resonant branches appear more clearly in blue, while the resonant  branches are evident as more diffuse bands due to dissipation below or above (depending on the quadrant) the non-resonant branches. Furthermore, Figure \ref{dispersion}b portrays the dispersion for the perpendicular waves for the same case by Fourier-transforming the perpendicular direction, showing $\omega$ vs. $k_y$. The dispersion of the oblique wave power is evident with most of the power concentrated at low-$k_y$ region, consistent with the 2D power spectrum shown in Figure \ref{2dpowerspec}. The diminished power at higher $k_y$ is signature of dissipation of the oblique waves. Such branches are not obvious in models of the expanding solar wind with homogenous background (e.g. \cite{MOV14}).}

\section{Discussion and Conclusions}
\label{conc:sec}
\begin{figure}
\noindent\includegraphics[scale=0.5]{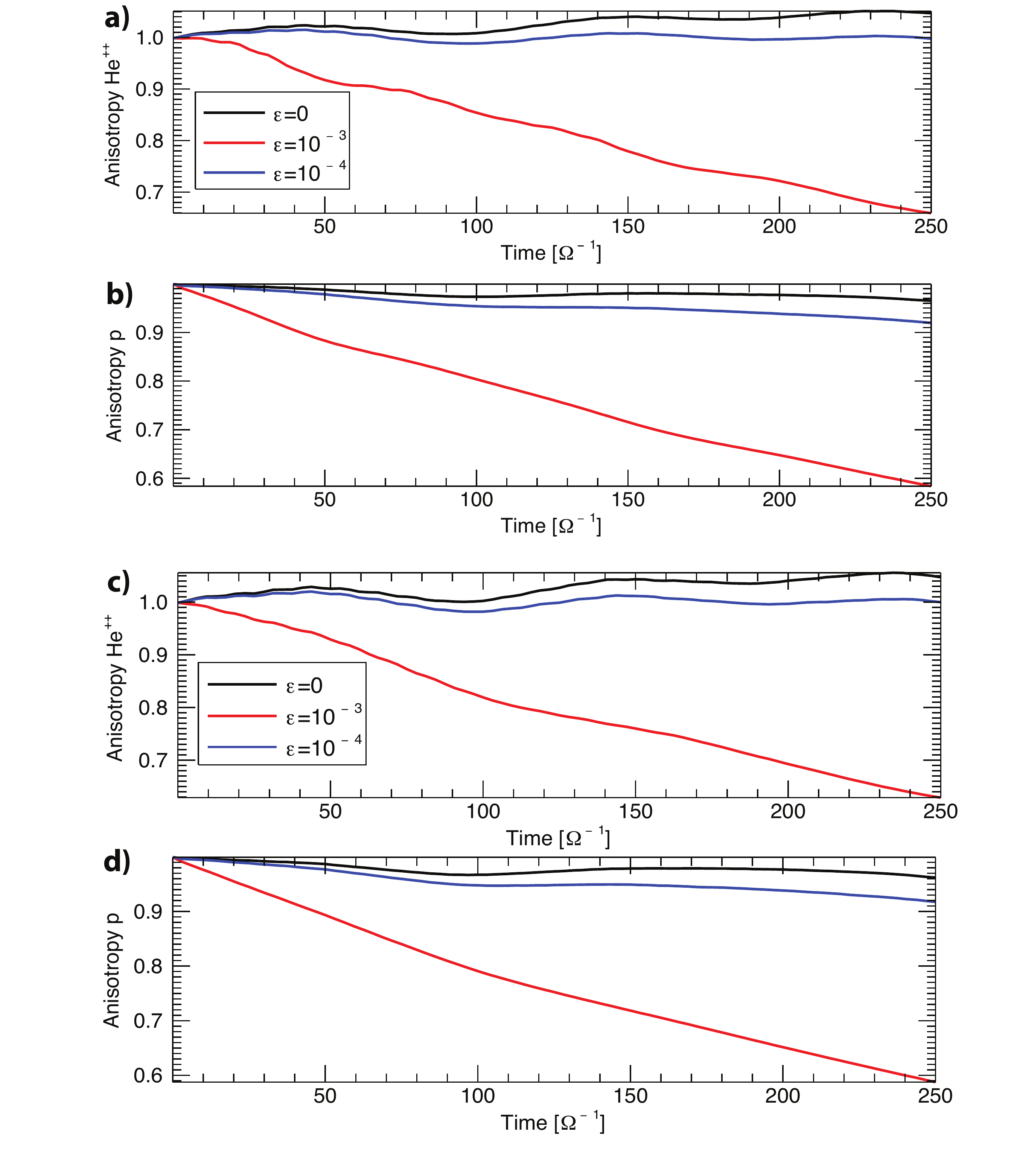}
 \caption{Temporal evolution of He$^{++}$ and proton temperature anisotropies for an initial sub-Alfv\'enic drift ($V_{d}=0.5V_{A}$). Case 1 with a Gaussian inhomogeneity ($q=2$) and Case 2 with a sharp inhomogeneity ($q=6$) are shown in the top (a and b) and bottom (c and d), respectively.}\label{subalfv_gauss}
 \end{figure}

\begin{figure}
\noindent\includegraphics[scale=0.6]{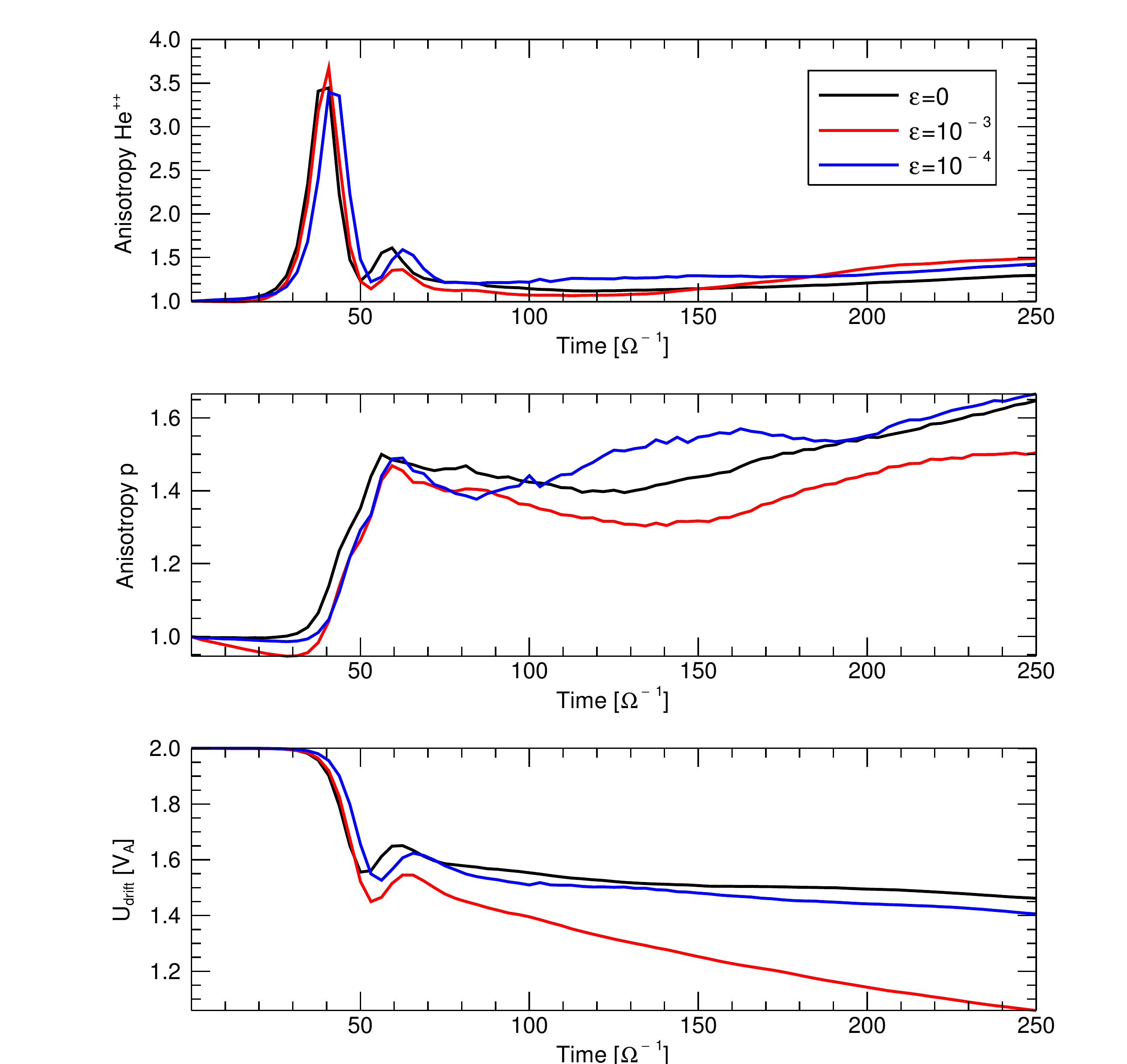}
 \caption{Temporal evolution of He$^{++}$ and proton temperature anisotropies for an initial super-Alfv\'enic drift ($V_{d}=2V_{A}$) an initial Gaussian inhomogeneity ($q=2$), \textbf{Cases 3 a--c}, for non-expanding (black) and expanding solar wind. } \label{superalfv_gauss}
 \end{figure}

\begin{figure}
\noindent\includegraphics[scale=0.75]{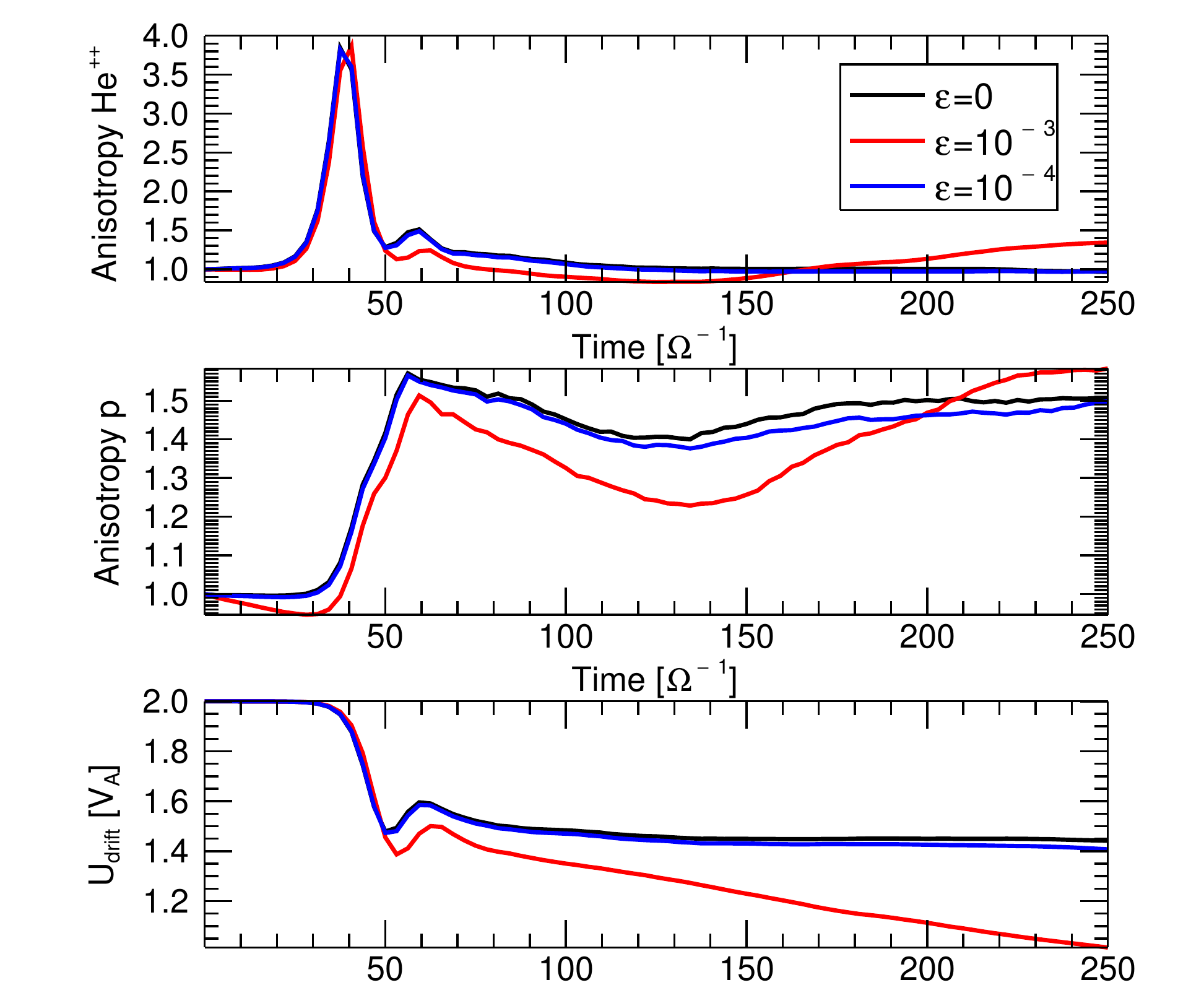}
\caption{Temporal evolution of He$^{++}$ and proton temperature anisotropies for an initial super-Alfv\'enic drift ($V_{d}=2V_{A}$) an initial sharp inhomogeneity ($q=6$), \textbf{Cases 4 a--c}, for non-expanding (black) and expanding solar wind. } \label{superalfv_sharp}
\end{figure}

\begin{figure}
\begin{tabular}{cc}
q=2 & q=6\\
\noindent\includegraphics[scale=0.4]{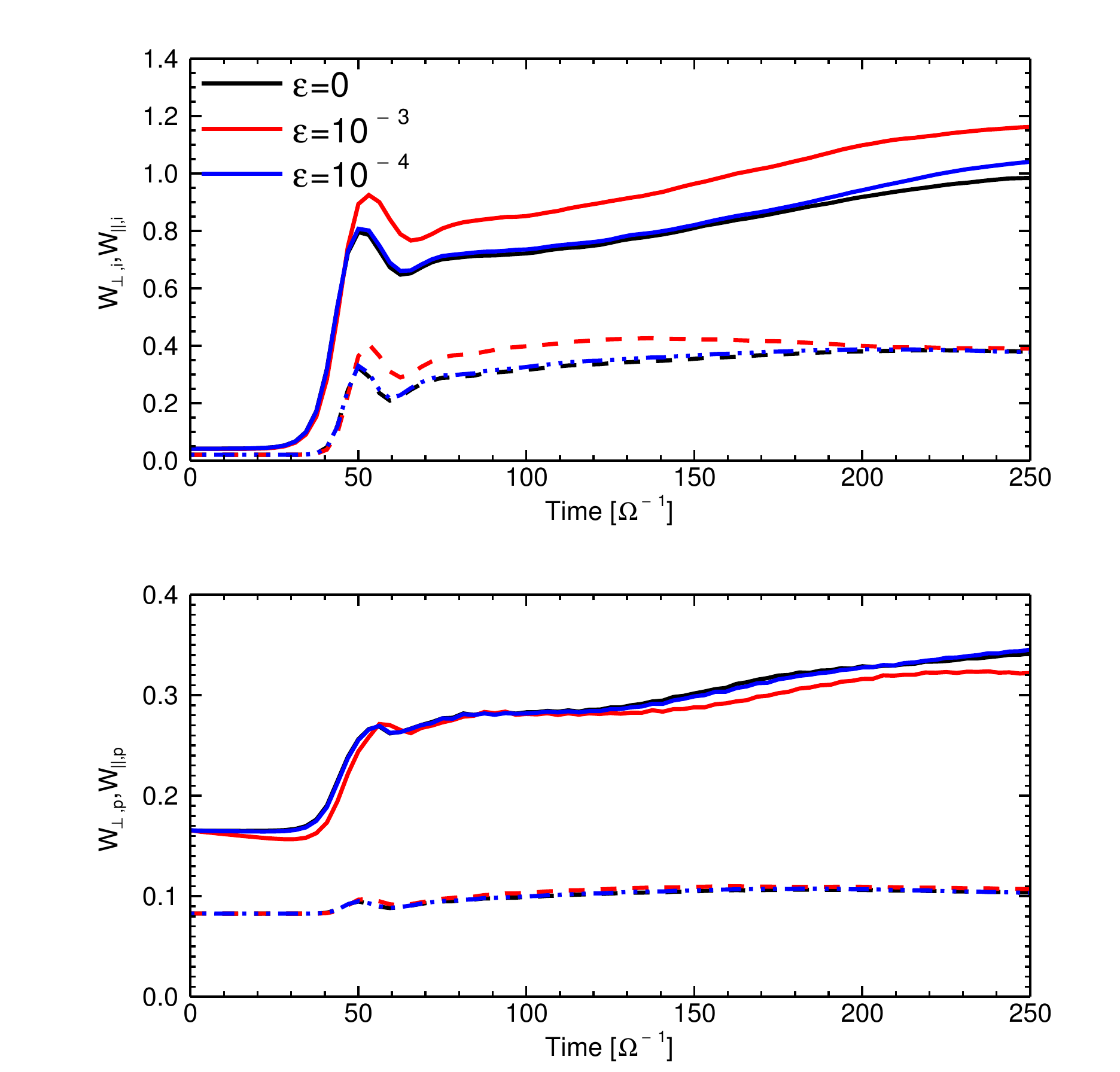}&
\noindent\includegraphics[scale=0.4]{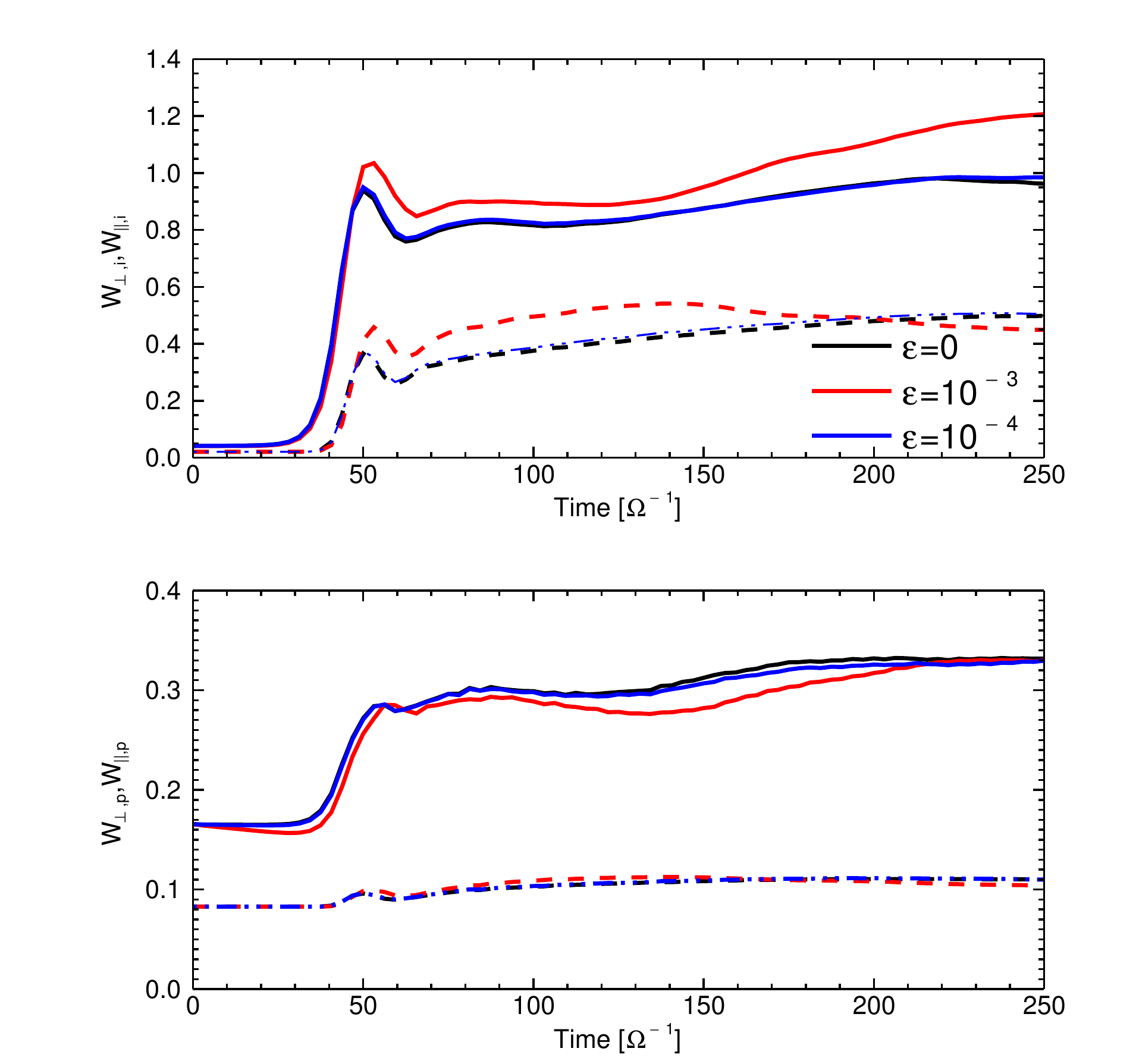}\\
\end{tabular}
\caption{Perpendicular and parallel \textbf{kinetic} energies for protons and ions in solar wind with initial super-Alfv\'enic drift ($V_{d}=2 V_{A}$) with a Gaussian inhomogeneity on the left column ($q=2$, \textbf{Case 3}) and a sharp inhomogeneity on the right column ($q=6$, \textbf{Case 4}) .} \label{energies_exp_comp}
\end{figure}

We investigate the ion heating processes in inhomogeneous expanding proton-He$^{++}$ solar wind plasma close to the Sun that may lead to temperature anisotropies observed in the fast solar wind streams. We analyze the effect of the relative ion drift instabilities and the turbulent Alfv\'en wave spectrum injected at the boundary  on inhomogeneous background density plasma on the heating of the expanding solar wind. We find that \textbf{for cases with super-Alfv\'enic relative drifts} the expansion of the solar wind has a small effect on the temperature anisotropy of the He$^{++}$ ions \textbf{(see Figures \ref{superalfv_gauss} and \ref{superalfv_sharp})}, which is reduced by the inhomogeneous plasma background and increased by resonant waves' perpendicular heating. On the other hand, the expansion has a more significant effect on the proton temperature anisotropy due to the smaller (compared to He$^{++}$ ions) perpendicular heating by the instability or the injected waves. \textbf{Furthermore, the expansion has a significant cooling effect for both ions and protons in cases where sub-Alfv\'enic drifts were included (see Figure \ref{subalfv_gauss})}. When the temperature anisotropy and heating of the plasma are due to drift instability or the injection of a turbulence wave spectrum, the heating dominates the evolution of the temperature anisotropy and the adiabatic cooling from the expansion \textbf{for super-Alfv\'enic drifts} is insignificant. However, when the drift is sub-Alfv\'enic ($V_d < V_A$), the inclusion of the expansion does decrease the temperature anisotropy of the plasma significantly within the simulated time and more so of the protons. In this case there is no heating source that may have countered the cooling due to the expansion, and the effect of the expansion dominates. The He$^{++}$ ion anisotropy decreases by 10\% for slow expansion rate and up to 40\% for a moderate expansion rate and similarly for the proton temperature anisotropy. The ion drift is not significantly affected by the expansion, while in the super-Alfv\'enic case the drift damps faster, when the expansion is included.

As in previous hybrid modeling studies \citep[e.g.][]{Ofm10a,OVM11,OVM14}, we see that the presence of a super-Alfv\'enic drift leads to \textbf{magnetosonic} instability that affects the ions and to a rapid He$^{++}$ ion heating in the perpendicular direction increasing the ion temperature anisotropy. This anisotropy decays quickly due to secondary ion-cyclotron instability resulting in emission of secondary ion-cyclotron waves, which in turn resonate with the protons causing the proton temperature anisotropy to increase following the relaxation of the He$^{++}$ temperature anisotropy.  

\begin{figure}
\begin{tabular}{cc}
\includegraphics[scale=0.45]{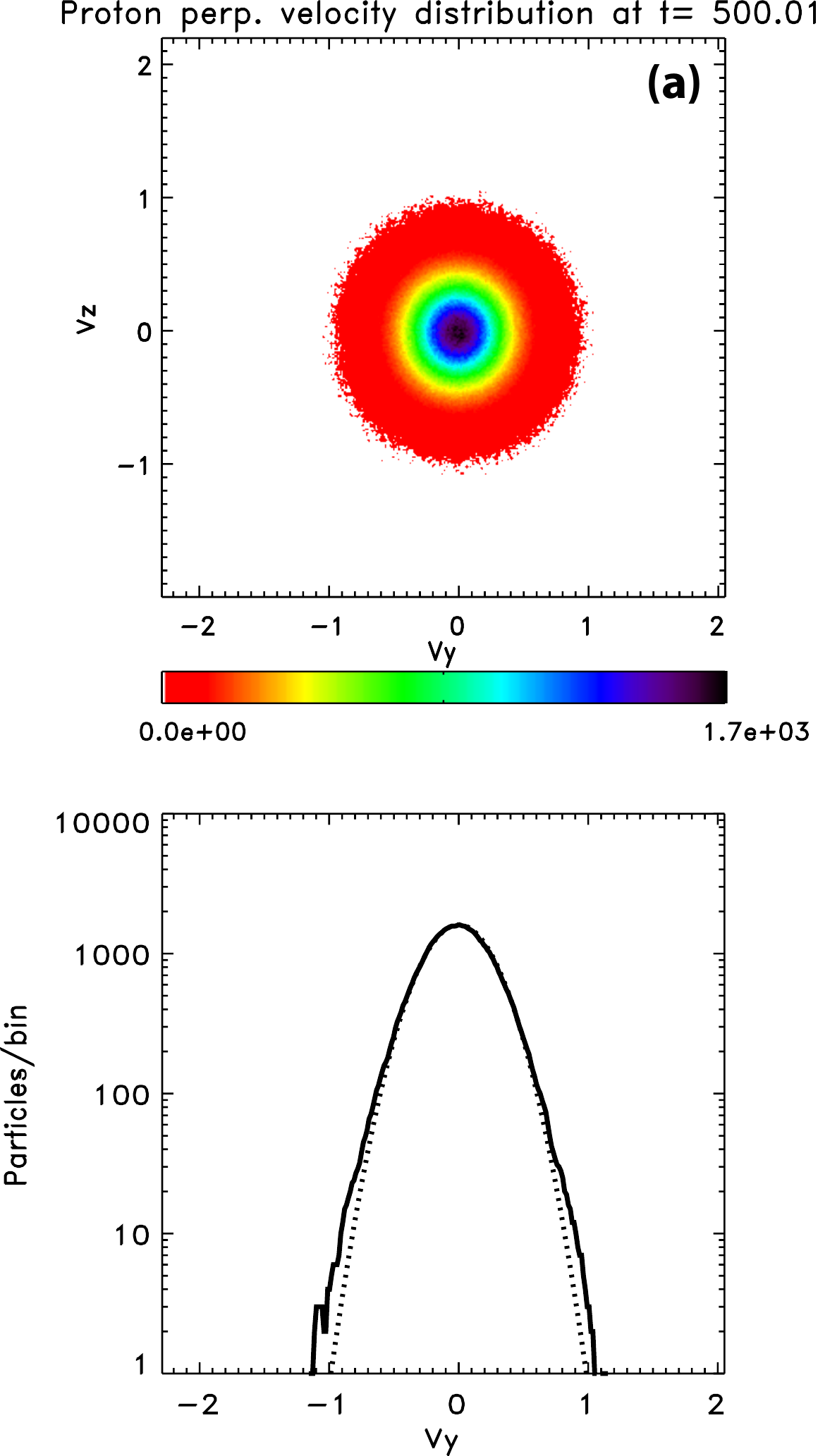} &
\includegraphics[scale=0.45]{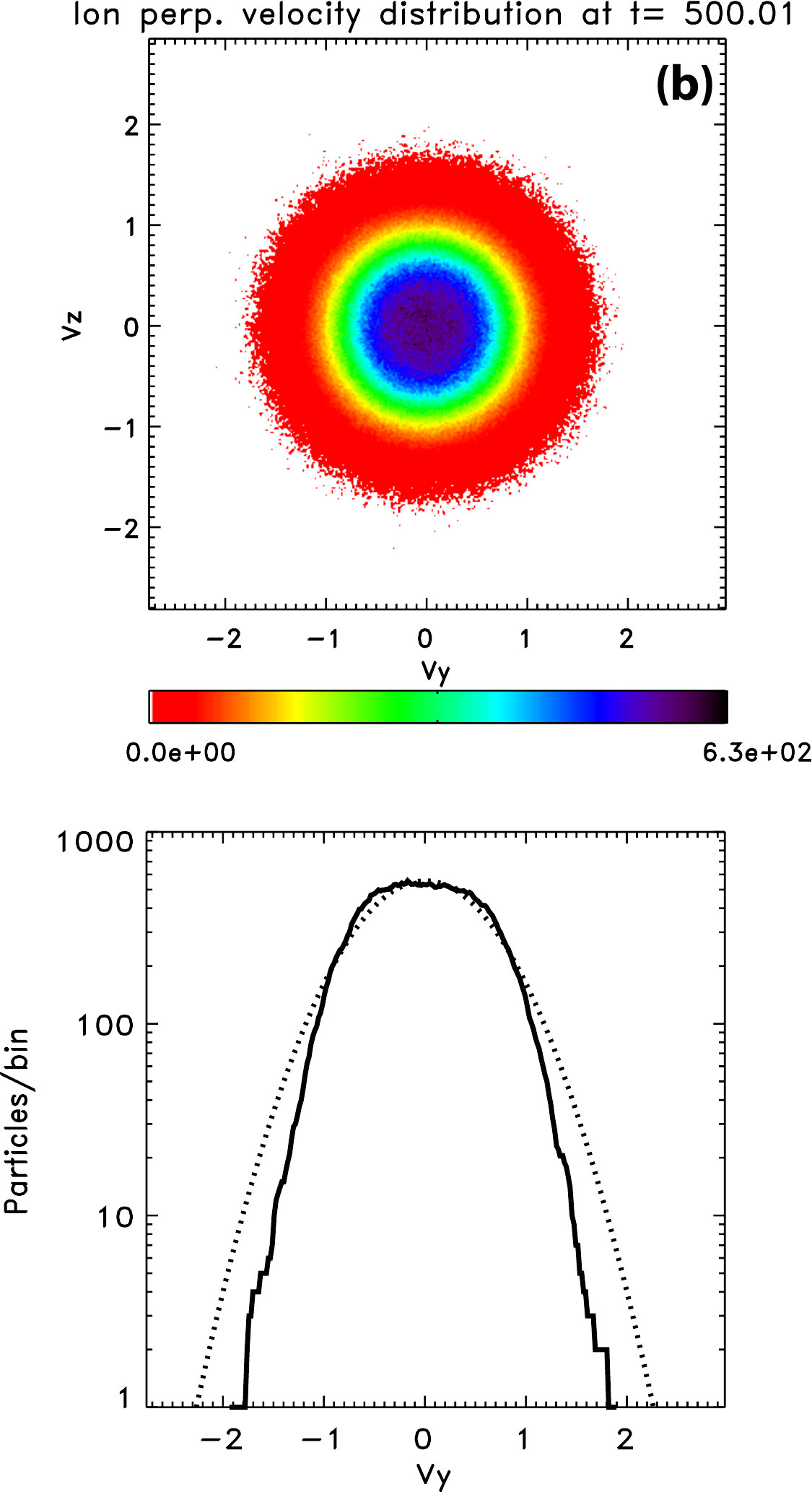}   \\
\includegraphics[scale=0.45]{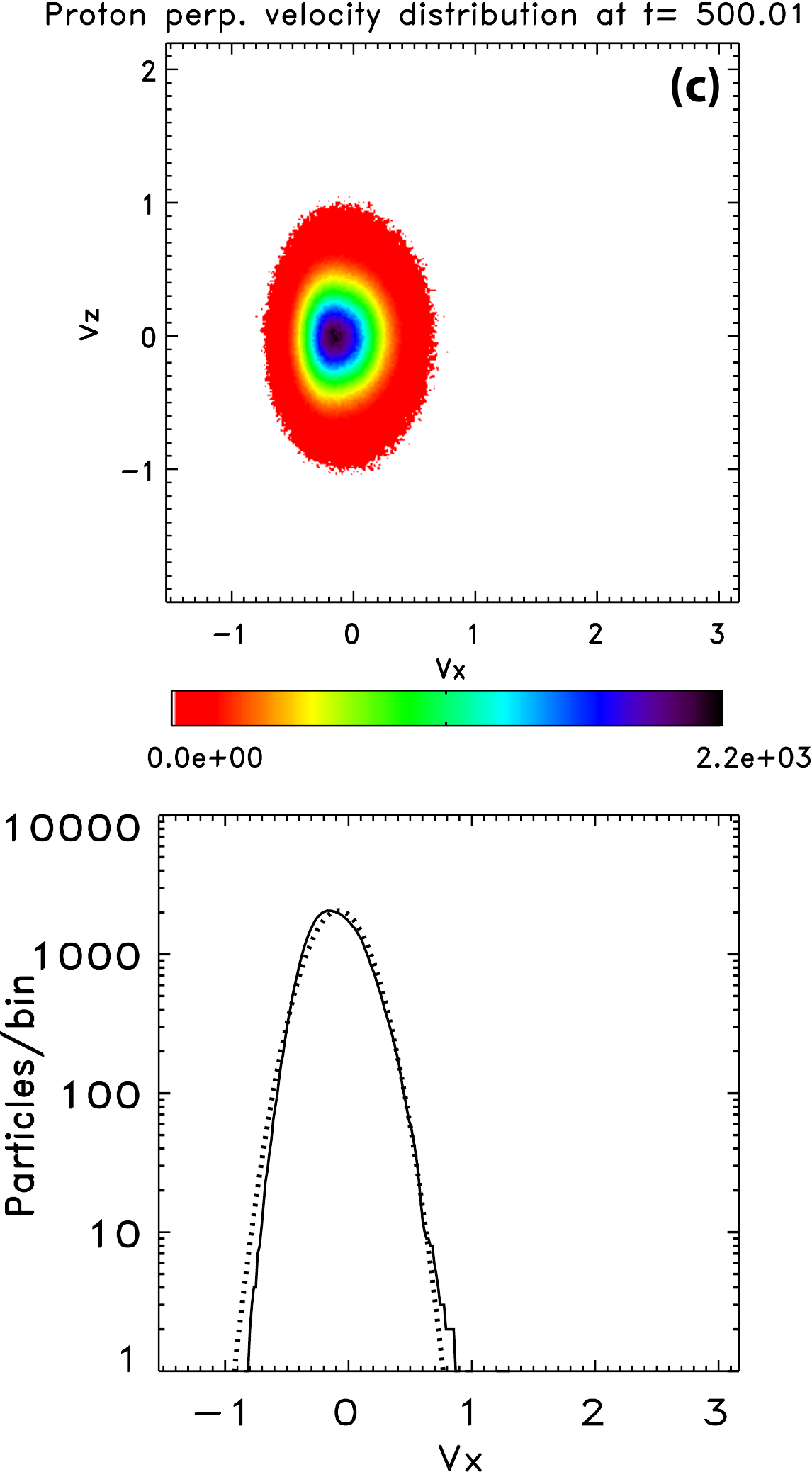}& 
\includegraphics[scale=0.45]{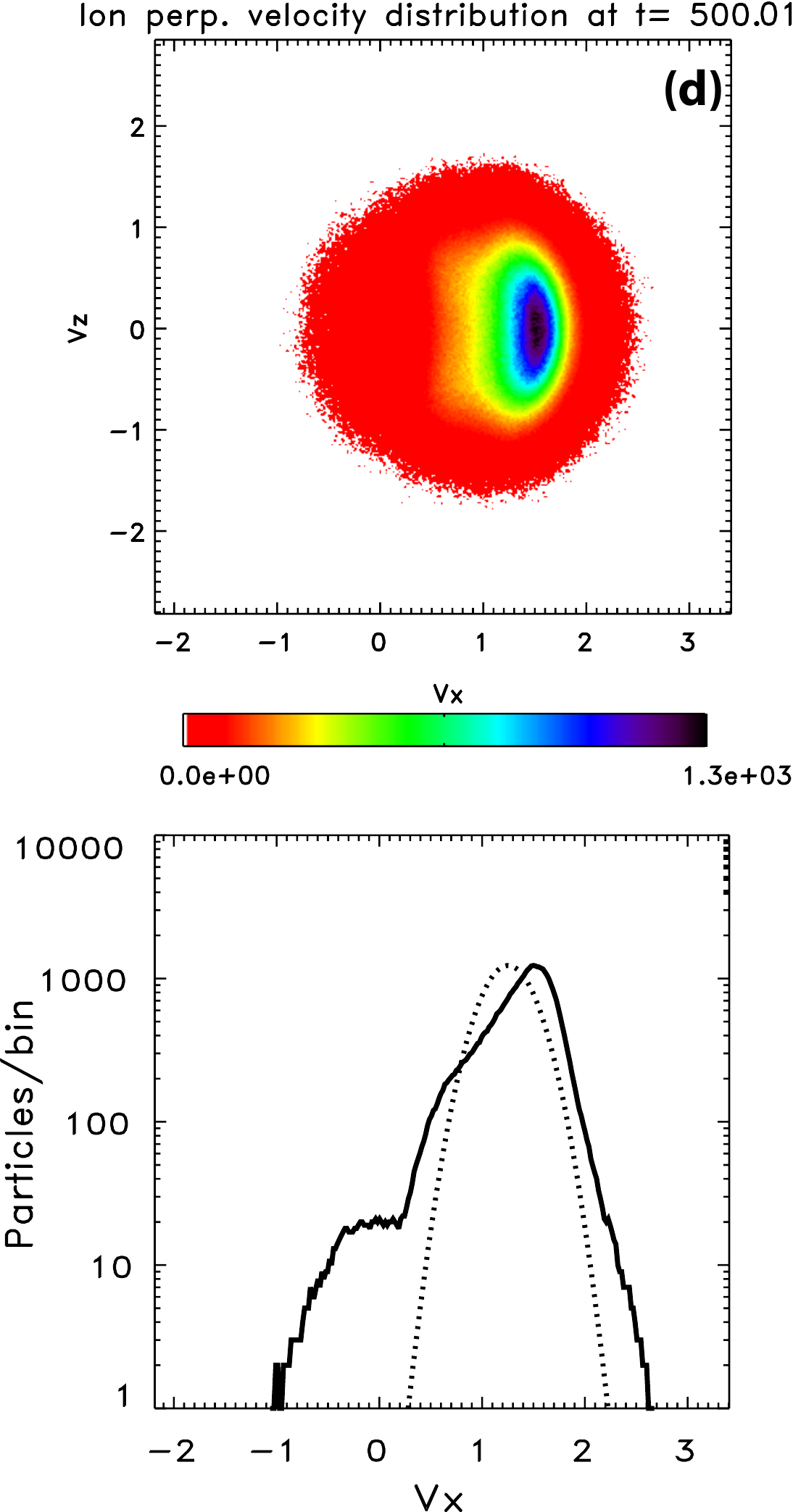}\\
\end{tabular}
\caption{A snapshot of the phase-space plots at $t = 500 ~\Omega^{-1}_p$ for a Gaussian inhomogeneity. All cases include a solar wind expansion rate $\epsilon=10^{-4}$ and an initial drift of $V_{d} = 2V_{A}$. Figures (a) and (b) show the proton and ion velocity distributions in the $V_y$ -- V$_z$ plane (\textbf{Case 3b}). Figures (c) and (d) show the proton and ion velocity distributions in the $V_x$ -- $V_z$ plane. The best-fit Maxwellian velocity distribution is shown with the dotted line.} \label{vspaceGauss}
\end{figure}

\begin{figure}
\begin{tabular}{ccc}
\includegraphics[scale=0.4]{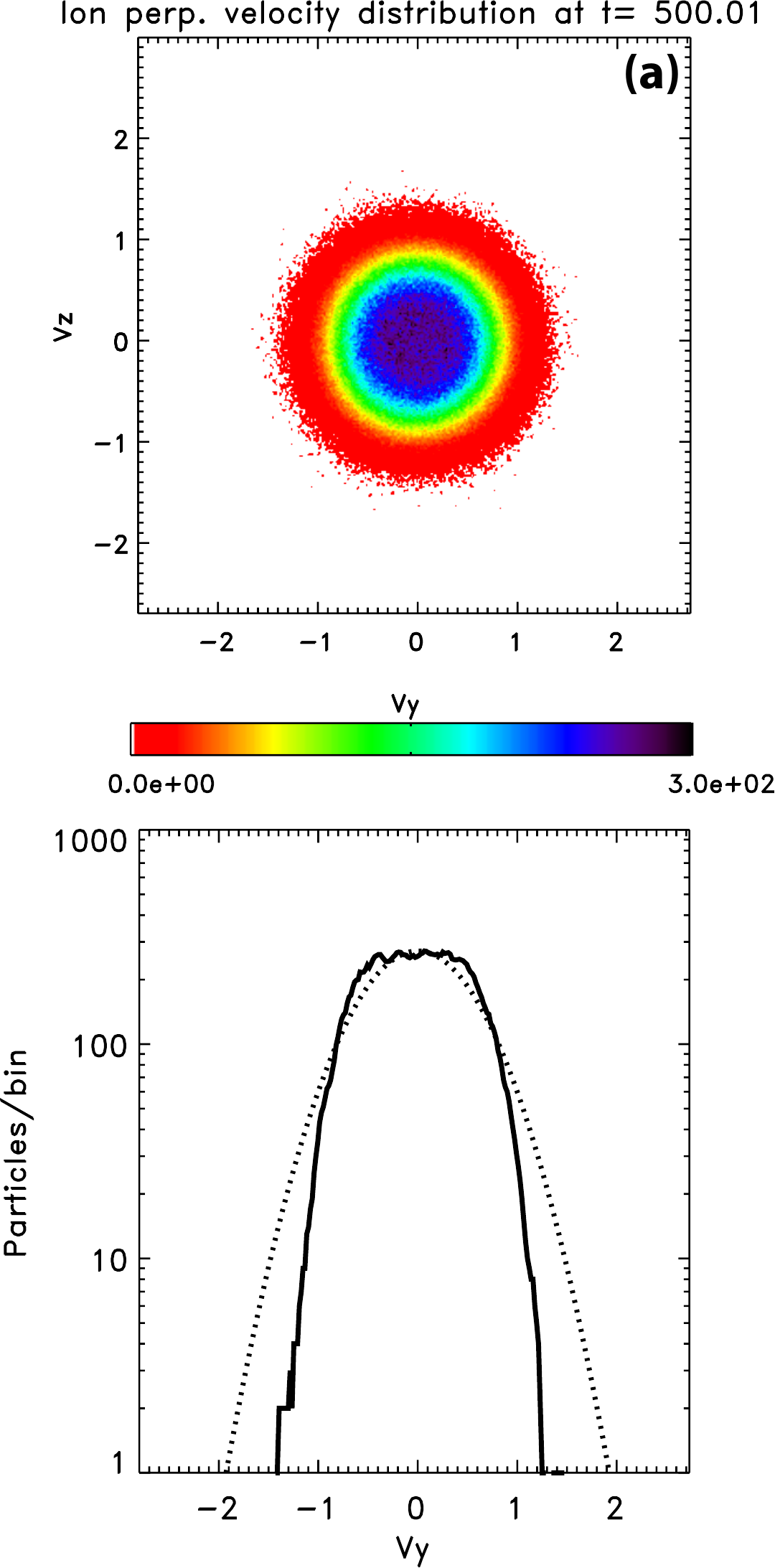}&
\includegraphics[scale=0.4]{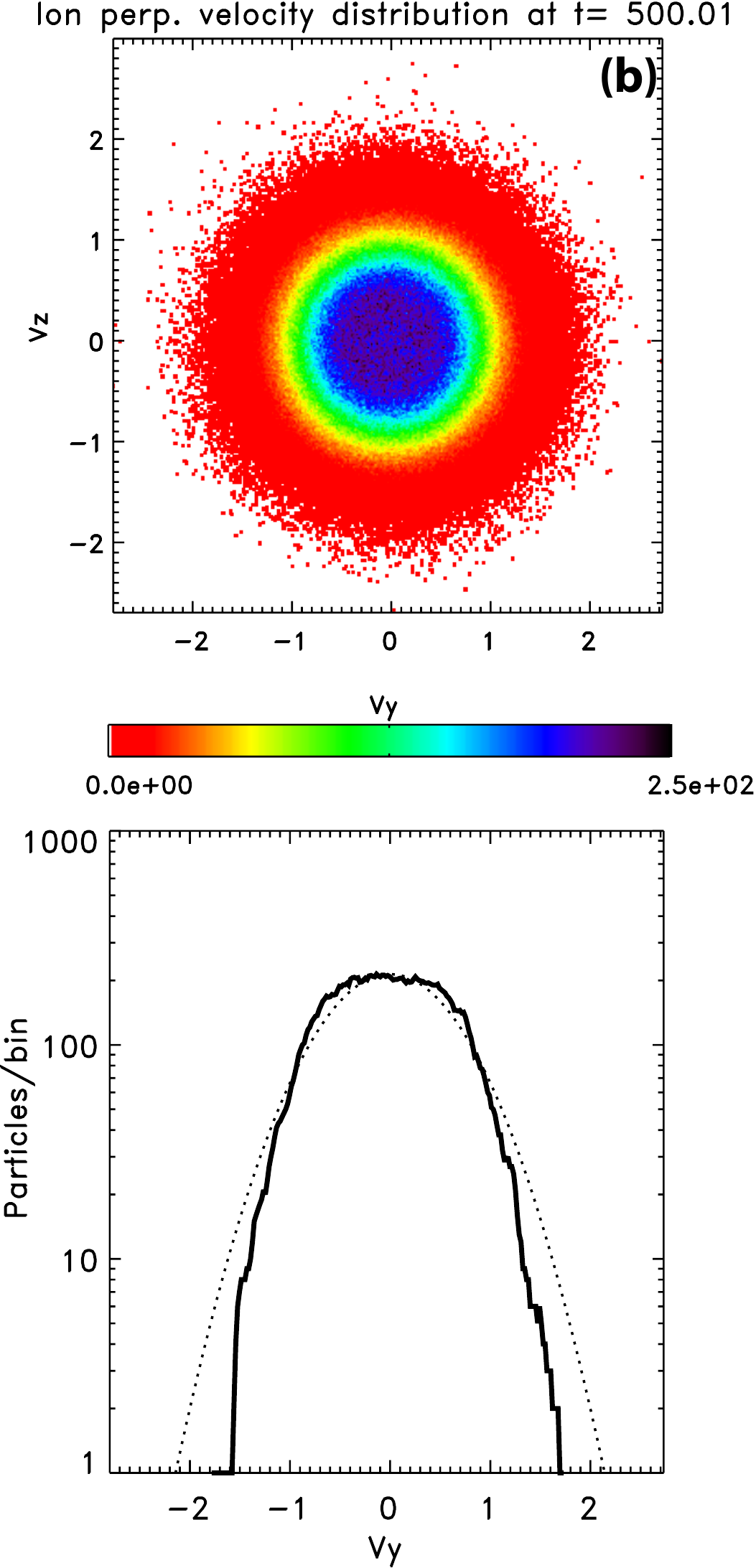} &
\includegraphics[scale=0.4]{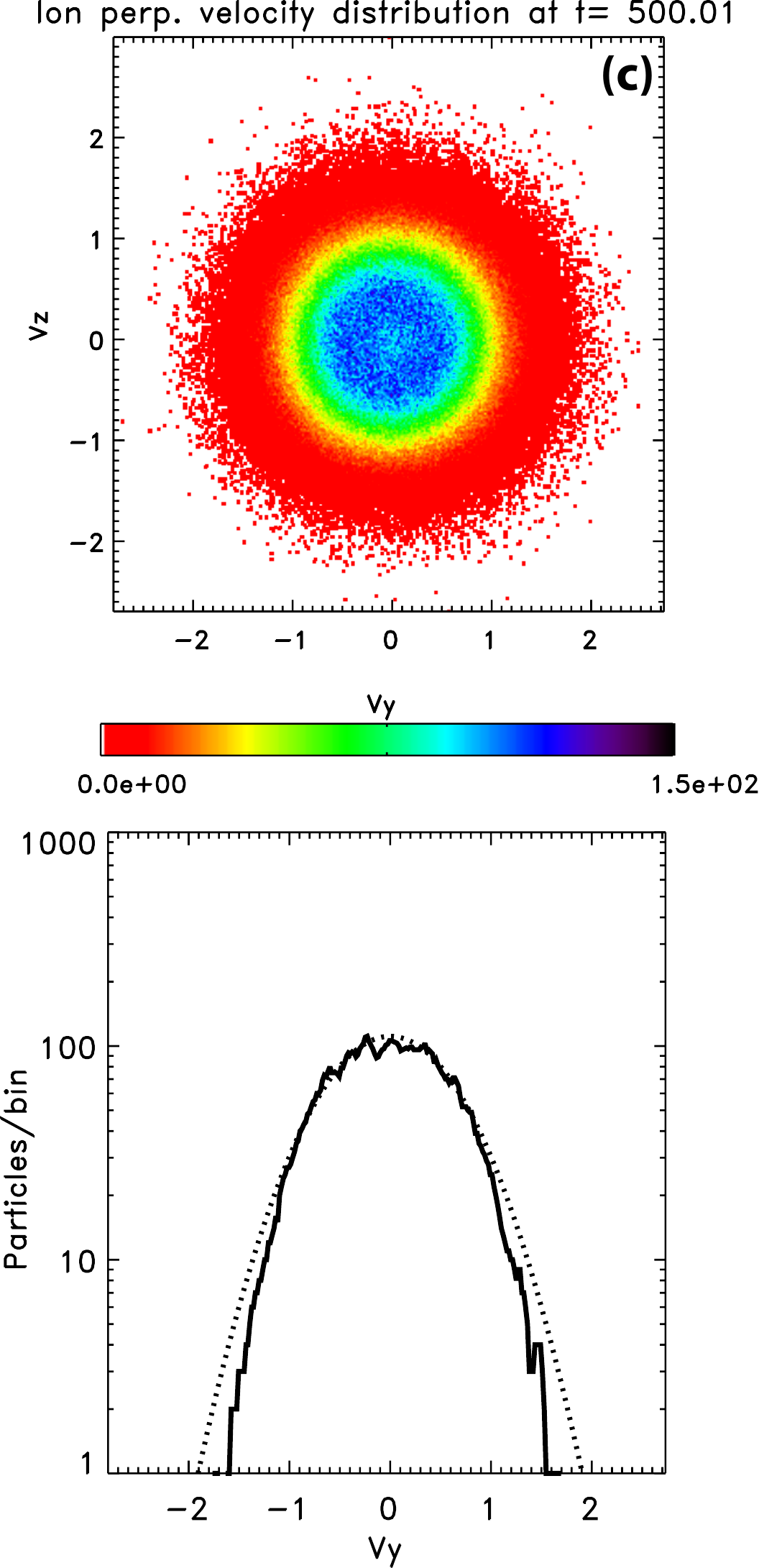}  \\
\end{tabular}
\caption{A snapshot of the $V_y$ -- $V_z$ phase space plot at $t = 500~\Omega^{-1}_p$ for a sharp inhomogeneity. All cases include a solar wind expansion rate $\epsilon=10^{-4}$ and an initial drift of $V_{d} = 2V_{A}$ (\textbf{Case 4b}). (a) corresponds to the outer homogenous region, region A in Figure \ref{initial_density}, (b) corresponds to the inhomogenous regions, region B in Figure \ref{initial_density}, and (c) corresponds to the inner homogeneous region, region C in Figure \ref{initial_density}. The best-fit Maxwellian velocity distribution is shown with the dotted line.}\label{yz-phasespace}
\end{figure}

\begin{figure}
\begin{tabular}{ccc}
\includegraphics[scale=0.4]{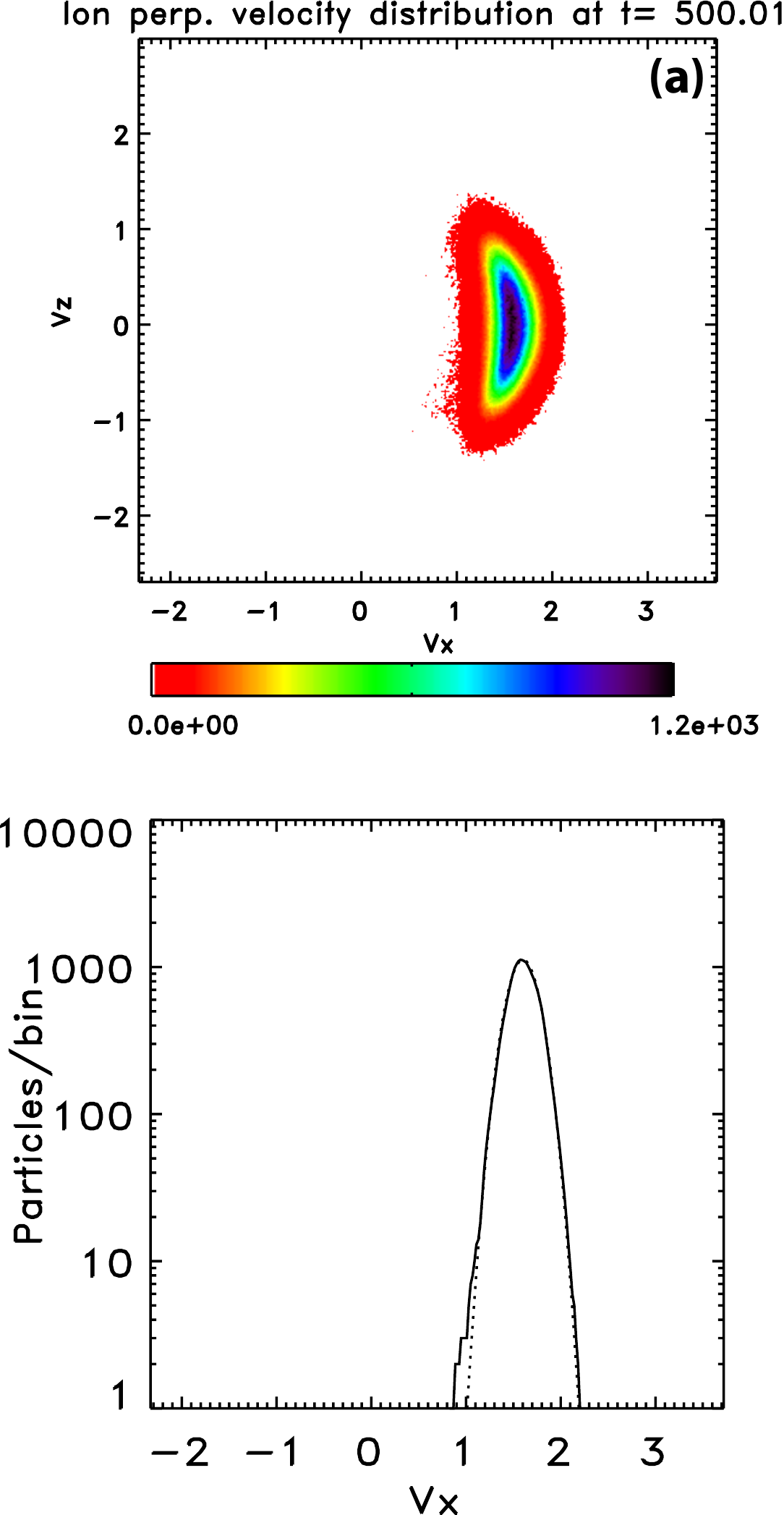}&
\includegraphics[scale=0.4]{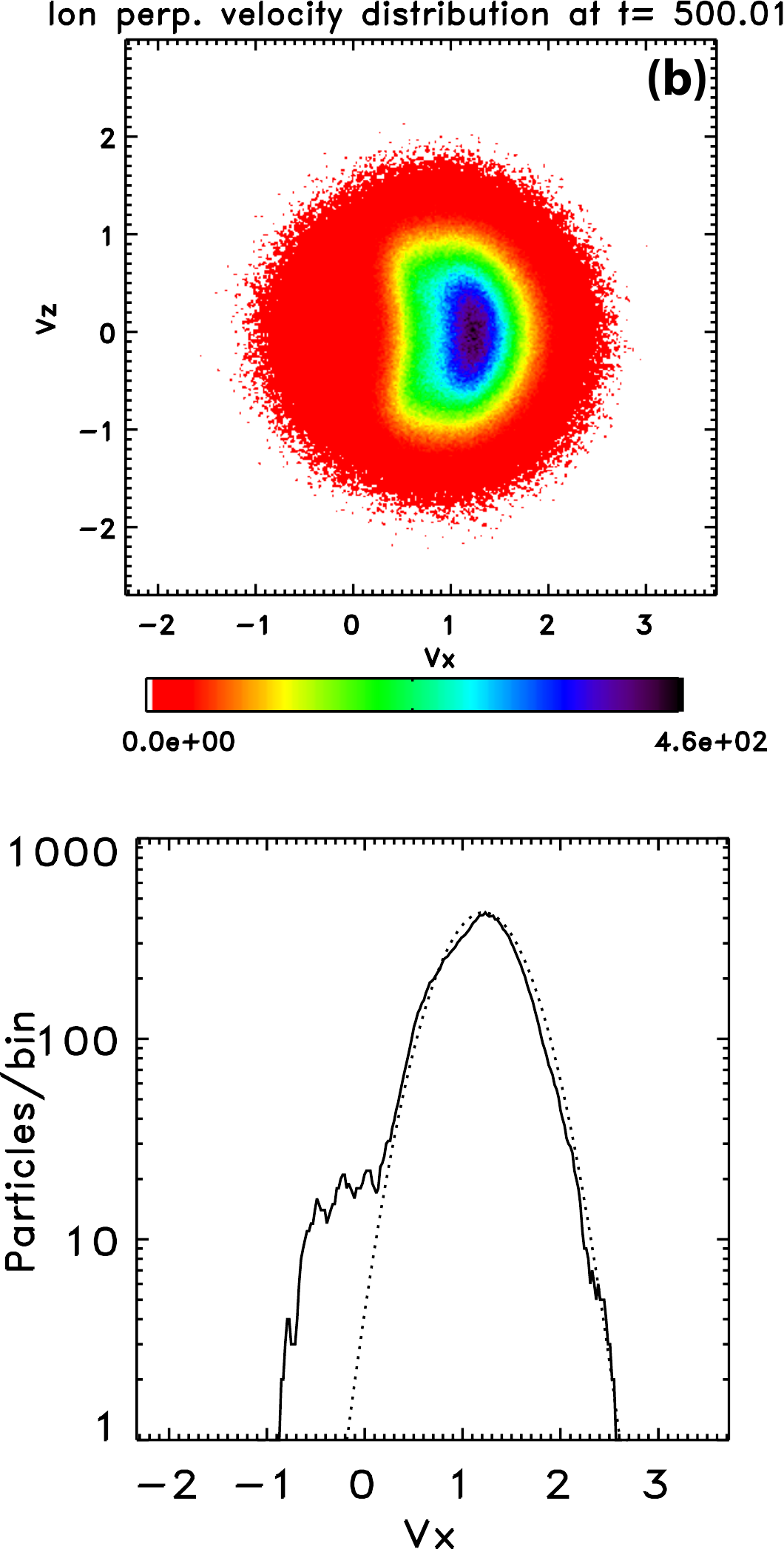}&
\includegraphics[scale=0.4]{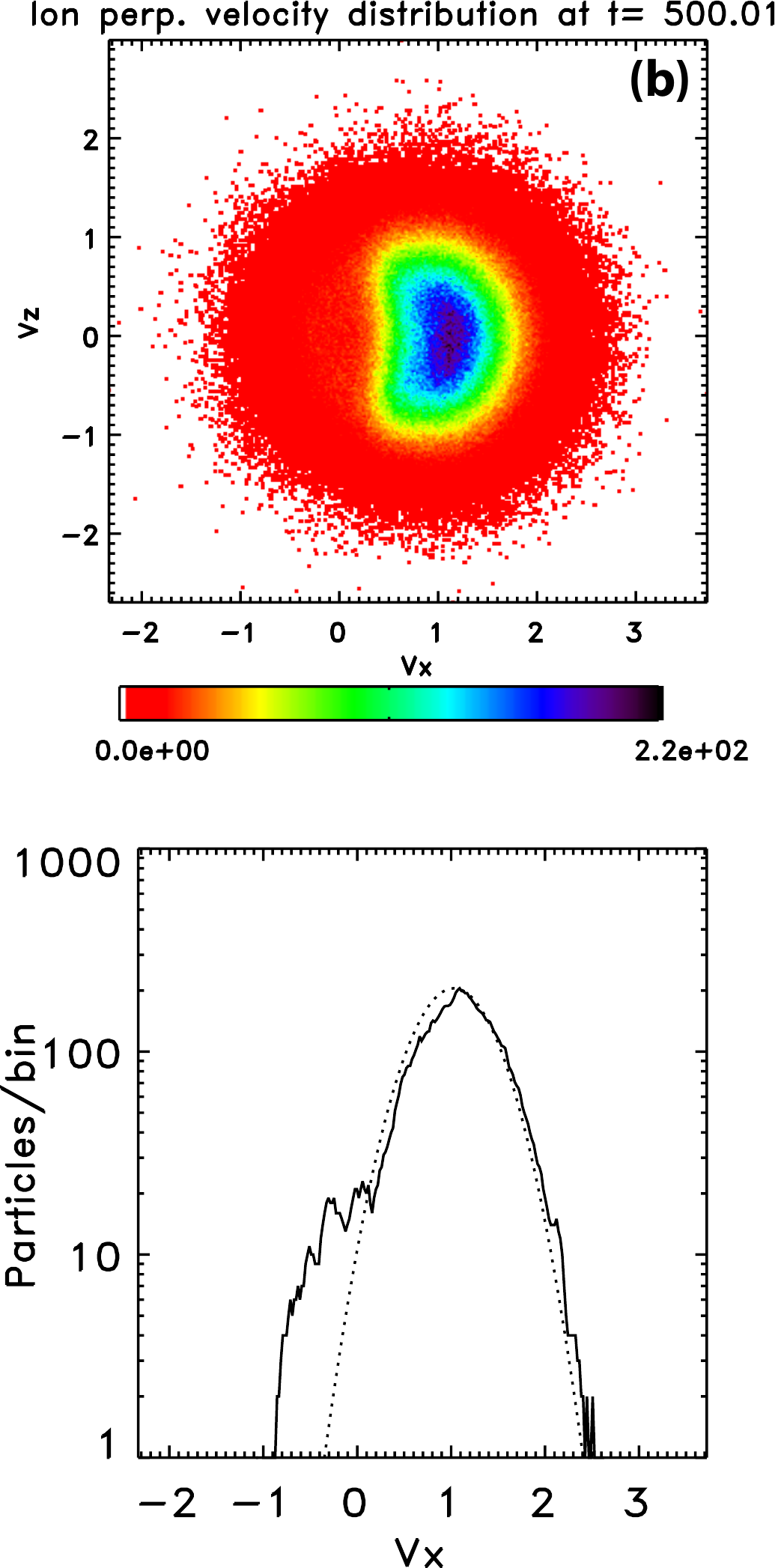}\\
\end{tabular}
\caption{A snapshot of the $V_x$ - V$_z$ phase-space plot at $t = 500  ~\Omega^{-1}_p$ for a sharp inhomogeneity.  All cases include a solar expansion rate $\epsilon=10^{-4}$ and an initial drift of $V_{d} = 2V_{A}$ (\textbf{Case 4b}). (a) corresponds to the outer homogenous region, region A in Figure \ref{initial_density} ,(b) corresponds to the inhomogenous regions, region B in Figure \ref{initial_density}, and (c) corresponds to the inner homogeneous region, region C in Figure \ref{initial_density}. The best-fit Maxwellian velocity distribution is shown with the dotted line.}
\label{xz-phasespace}
\end{figure}

\begin{figure}
\noindent\includegraphics[scale=0.7]{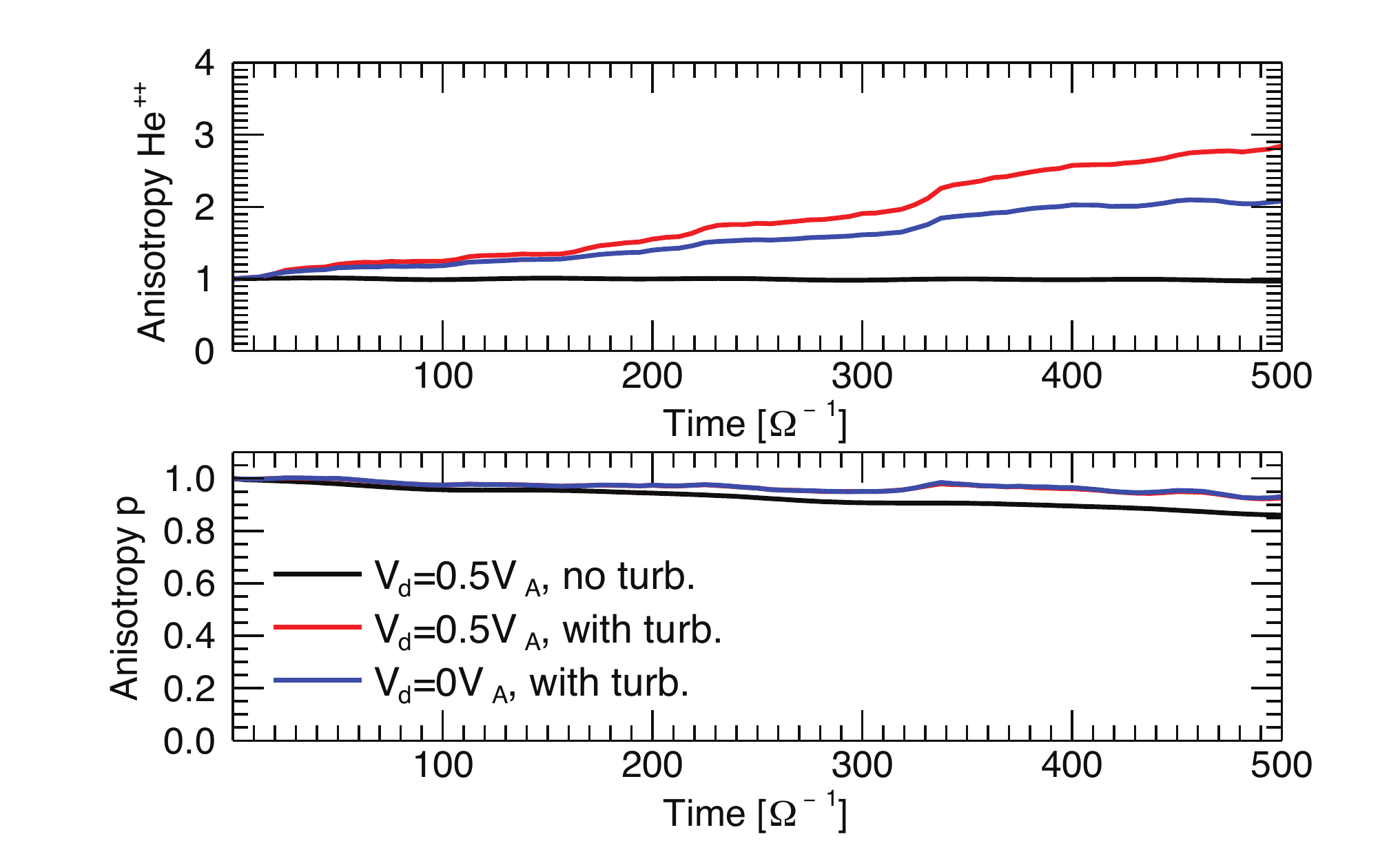}
\caption{Effect of turbulent wave spectrum for expanding solar wind ($\epsilon=10^{-4}$) on the temporal evolution of He$^{++}$ and proton temperature anisotropies for an initial sub-Alfv\'enic drift in a plasma with Gaussian inhomogeneity ($q=2$). The blue line shows the case with expansion and turbulent spectrum, but no drift \textbf{(Case 5b)}. The red line shows the case with expansion, turbulent spectrum and a drift of $V_{d}= 0.5 V_{A}$, \textbf{(Case 5c)}. The black line shows the case with drift and expansion, but no turbulent spectrum included \textbf{(Case 1b)}.}\label{subalfv_turb}
\end{figure}

\begin{figure}
\noindent\includegraphics[scale=0.7]{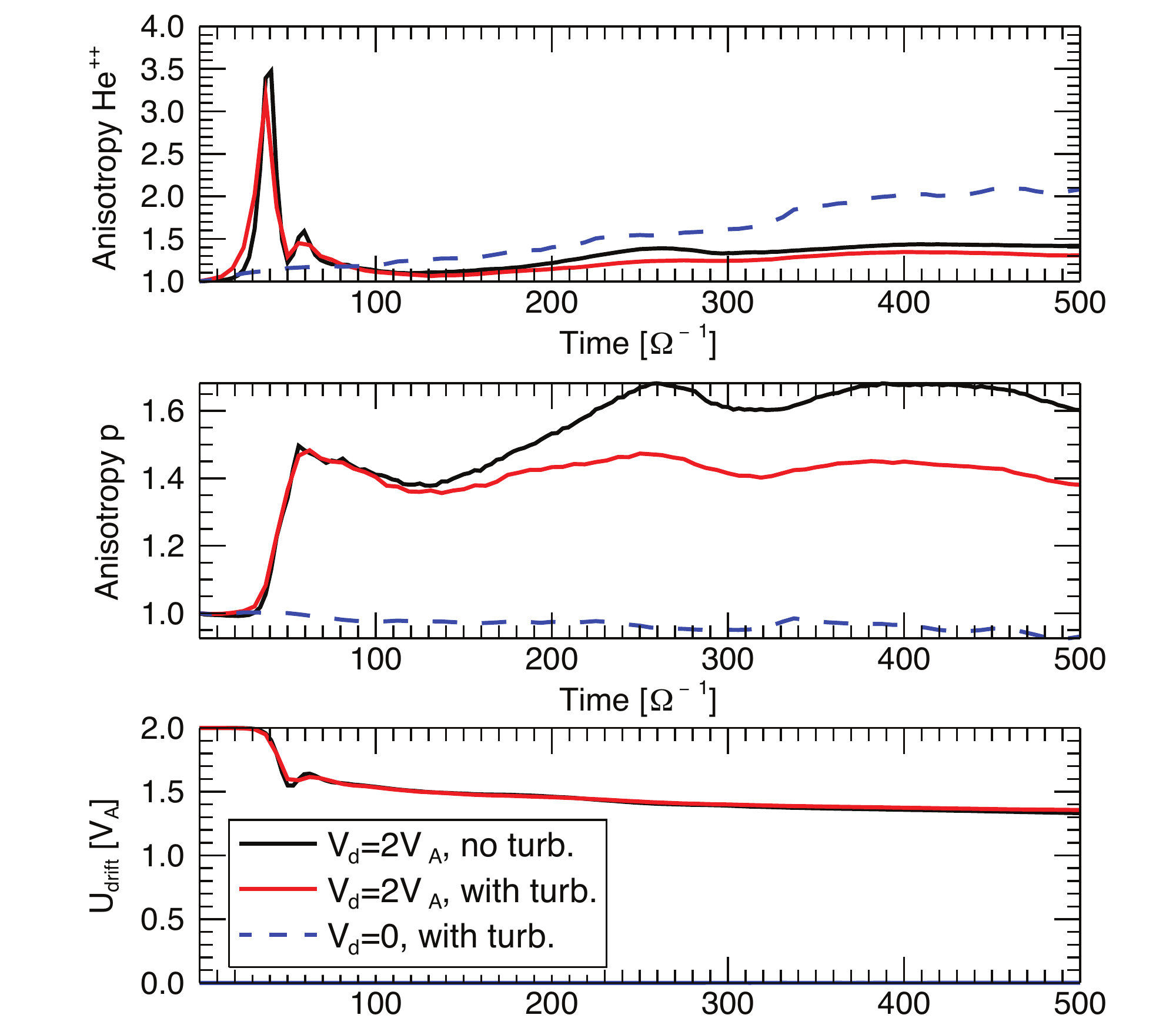}
\caption{Effect of turbulent wave spectrum for expanding solar wind ($\epsilon=10^{-4}$) on the temporal evolution of He$^{++}$ and proton temperature anisotropies for an initially super-Alfv\'enic drift ($V_{d}=2 V_{A}$) and a Gaussian inhomogeneity ($q=2$).  \textbf{The blue line shows the case with expansion and turbulent spectrum, but no drift (Case 5b). The red line shows the case with expansion, drift and turbulence (Case 5a). The black line shows the case with drift, expansion, but no turbulent spectrum (Case 3b).}}\label{superalfv_turb}
\end{figure}

\begin{figure}
\noindent\includegraphics[scale=0.7]{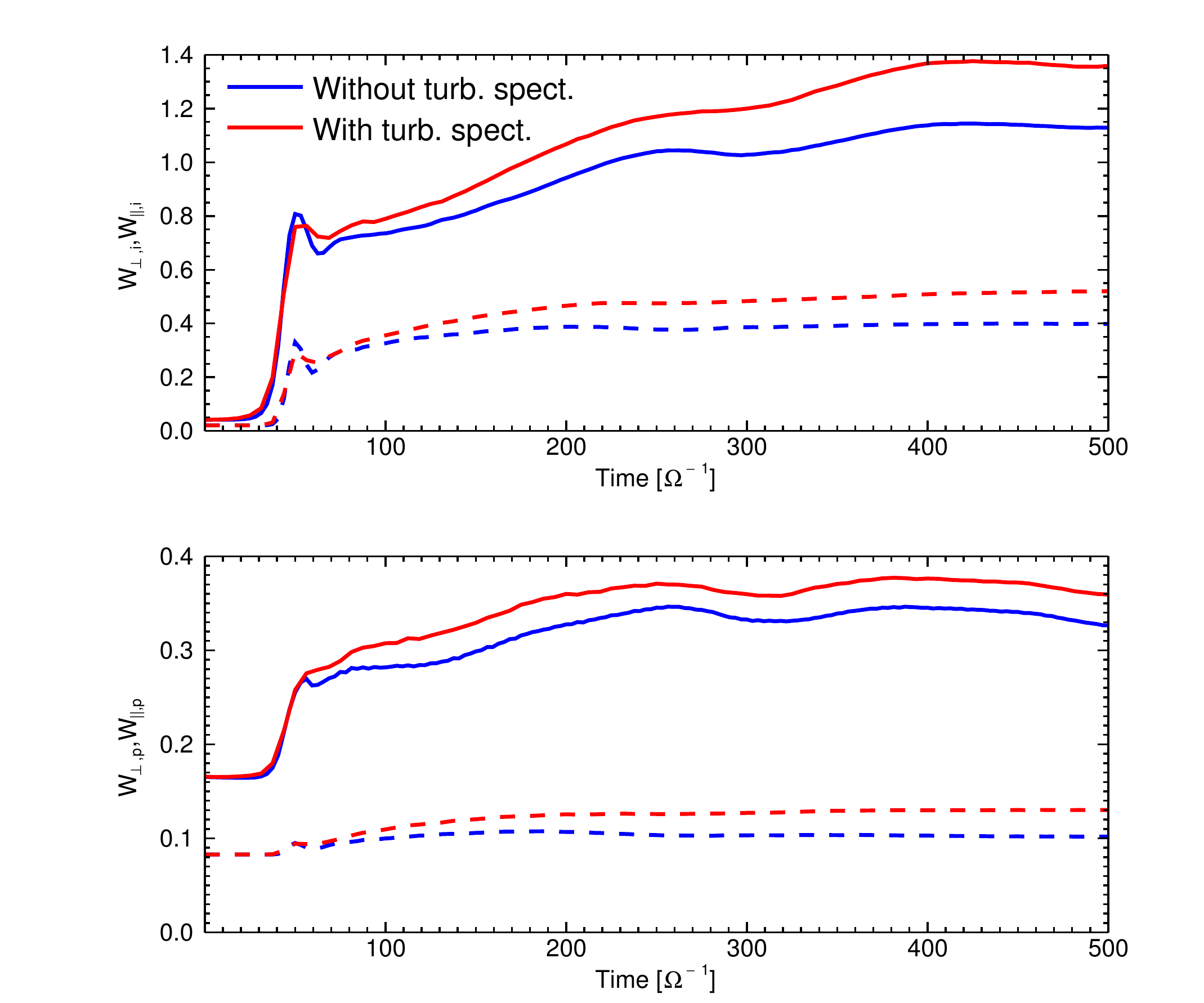}
\caption{Perpendicular and parallel \textbf{kinetic} energies for protons and ions in expanding solar wind ($\epsilon=10^{-4}$) with \textbf{(Case 5a)} and without \textbf{(Case 3b)} initial turbulent wave spectrum at the boundary and super-Alfv\'enic drift ($V_{d}=2 V_{A}$) with a Gaussian inhomogeneity ($q=2$).} \label{energies}
\end{figure}

\begin{figure}
\noindent\includegraphics[scale=0.7]{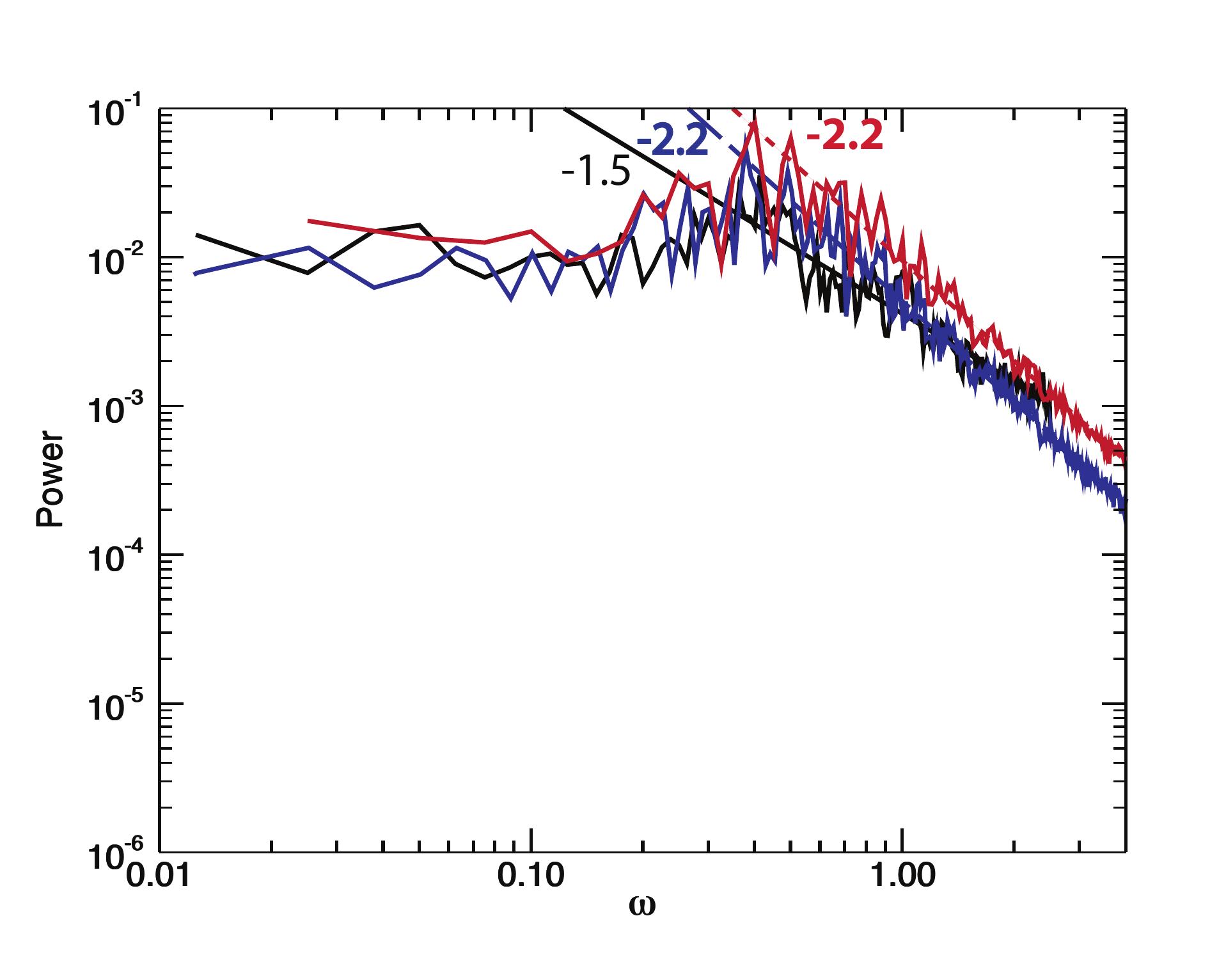}
\caption{Power spectrum of magnetic fluctuations for different cases explored in the inhomogenous plasma and corresponding best power law fit \textbf{at the end of the simulation ($\mathbf{t = 500  \Omega ^{-1}_p}$)}. The black line shows the results for Case 5a (slope = $-1.5$), \textbf{which includes the expansion and initial drift and turbulent spectrum}. The blue line shows the results for Case 3b (slope = $-2.2$), \textbf{which includes drift and expansion}. The red line shows the results for Case 3a (slope = $-2.2$), \textbf{which includes only an initial drift and no expansion}. The power law fit is shown with the dashed lines.} \label{powerspec_turb}
\end{figure}

\begin{figure}
\noindent\includegraphics[scale=0.7]{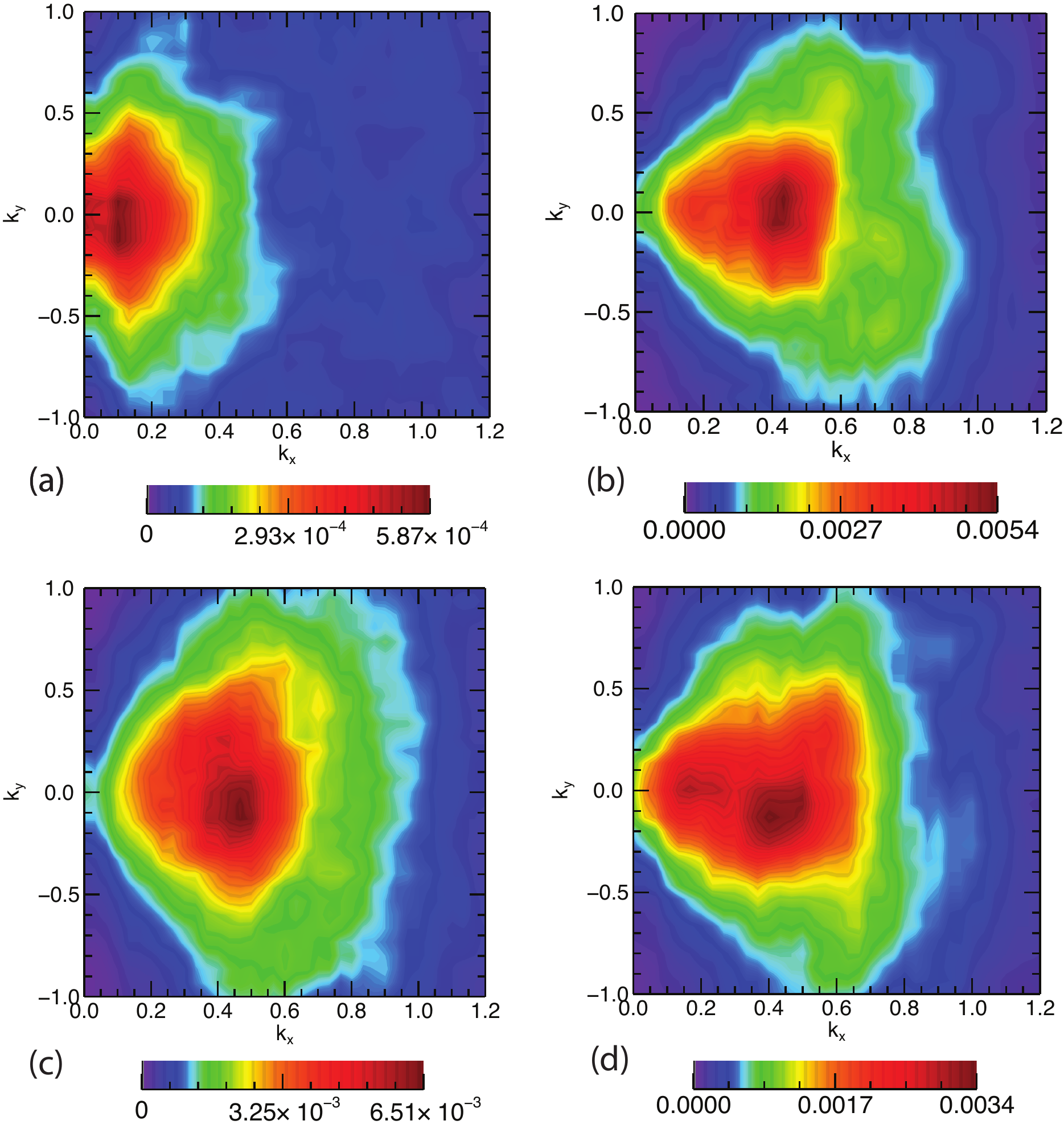}
\caption{Two-dimensional power spectrum for different conditions in solar wind plasma with a Gaussian inhomogeneity at t = 250 $\Omega^{-1}_p$: a) Only initial turbulent spectrum is considered, b) initial drift $V_{d}=2 V_{A}$ and no expansion or turbulence, c) initial drift ($V_{d}=2 V_{A}$) and solar wind expansion ($\epsilon = 10^{-4}$) included (Case 3b), d) initial drift and solar wind expansion included, as in c) plus initial turbulent spectrum included (Case 5a).}\label{2dpowerspec}
\end{figure}

\begin{figure}
\begin{tabular}{c}
\noindent\includegraphics[scale=0.6]{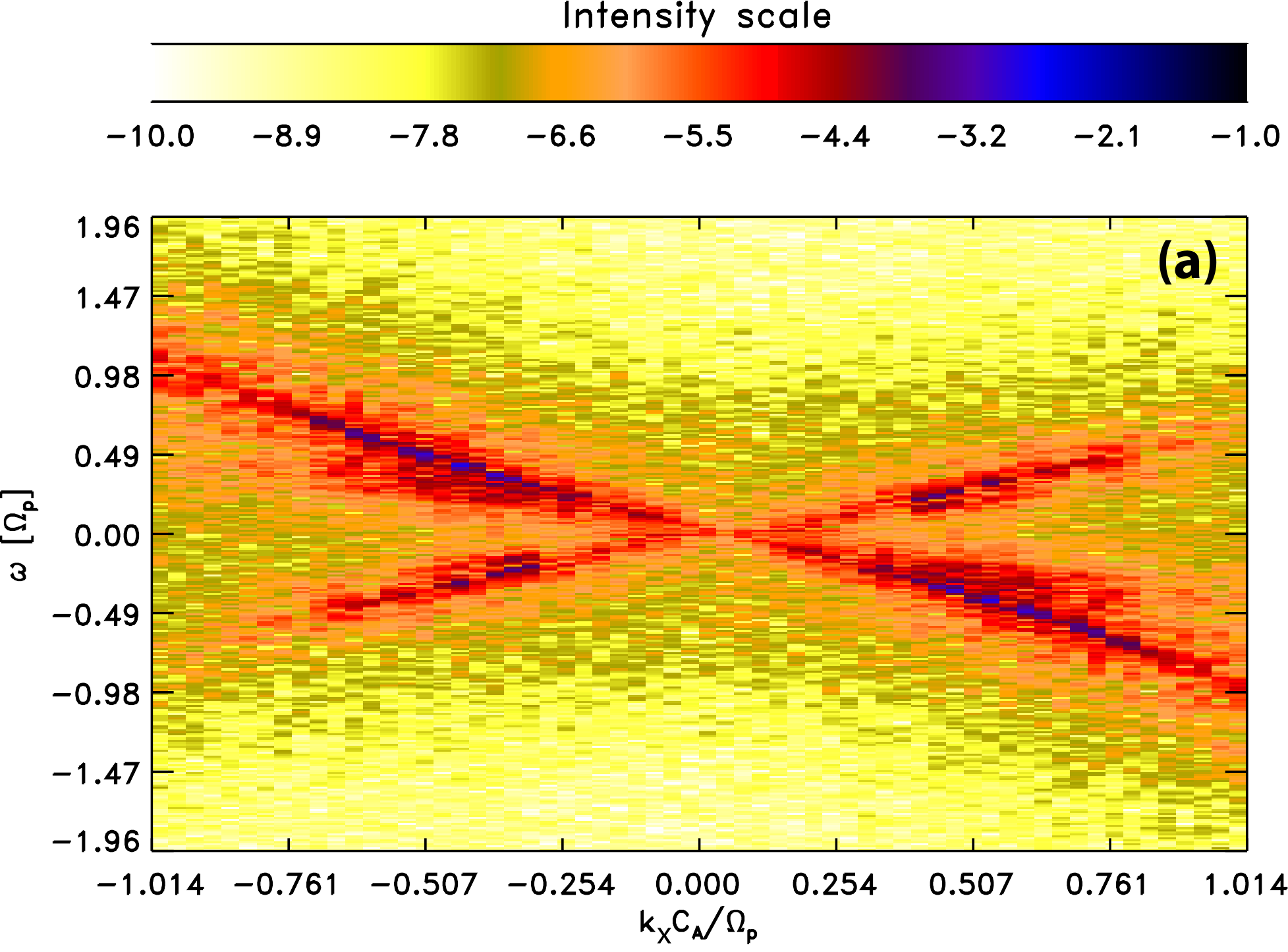}\\
\noindent\includegraphics[scale=0.6]{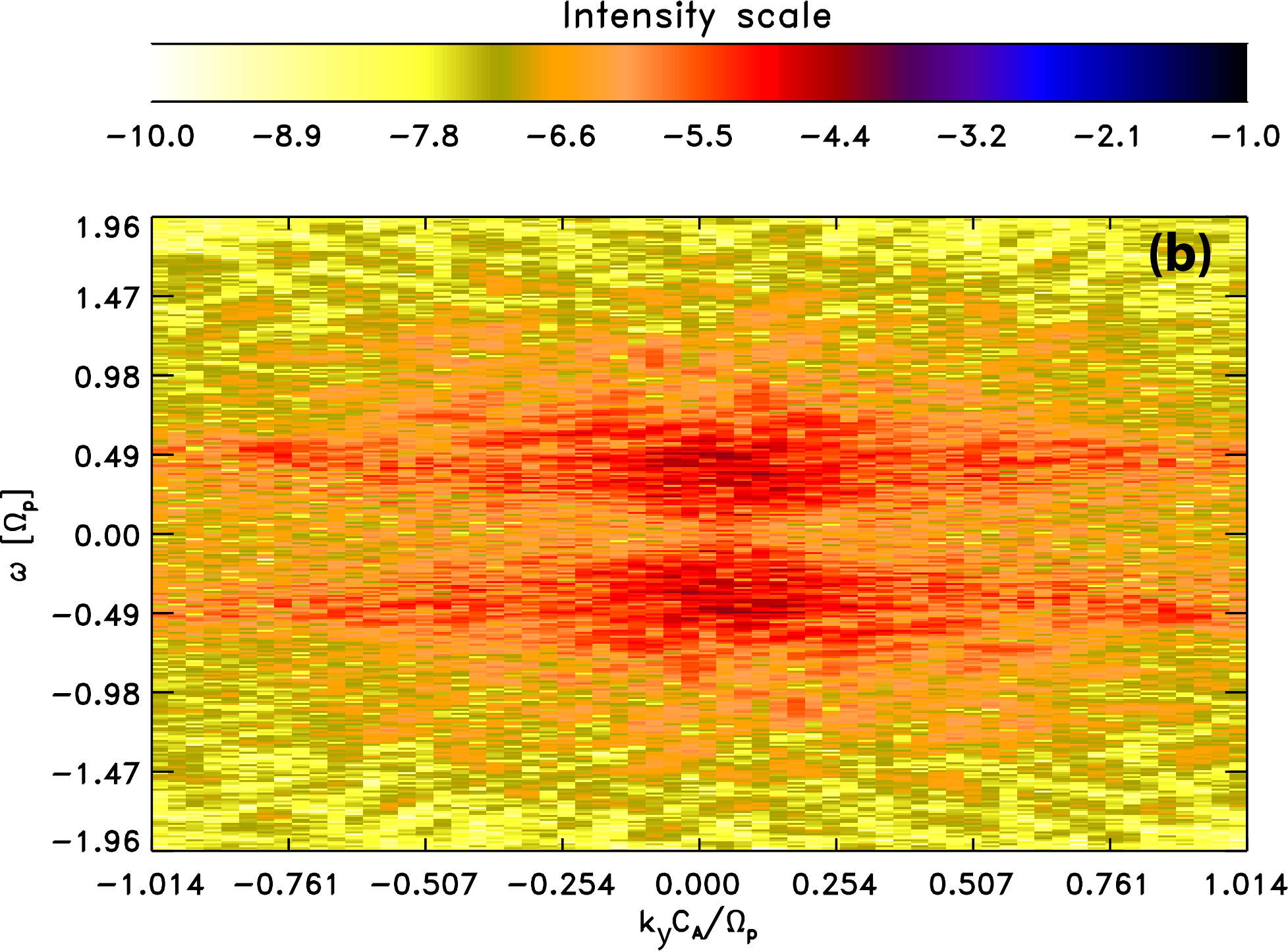}\\
\end{tabular}
\caption{Dispersion relation $\omega$ vs. $k$ obtained from the hybrid model simulation for super-Alfv\'enic drift in the inhomogenous expanding plasma (Case 3b). (a) shows the dispersion for the parallel wave number space ($\omega$ vs. $k_x$). (b) shows the dispersion for the perpendicular wave number space ($\omega$ vs. $k_y$), illustrating the oblique waves.}\label{dispersion}

\end{figure}

Our study includes an inhomogeneous density background in the initial state of the modeled solar wind plasma. The inhomogeneity does not dissipate on the time-scale considered in this study, since the low-$\beta$ plasma is near pressure-balance. We find that the inhomogeneity affects the time needed for the proton anisotropy to relax so that the sharper inhomogeneity case reaches a relaxed state faster than the Gaussian case, possibly due to the production of oblique waves that accelerate the evolution of the instability. The inhomogeneity also significantly affects the ion velocity distribution as seen in Figures~\ref{vspaceGauss}c, \ref{vspaceGauss}c, and  \ref{xz-phasespace}. We find that the velocity space diffusion is consistent with kinetic shell models \citep[e.g.][and references therein]{She04,Ise04}, and with previous hybrid modeling studies of drifting ions \citep[e.g.,][]{XOV04,OV07,OVM14}. The perpendicular heating produces a velocity distribution that is highly non-Maxwellian and which has a beam-like peak for the Gaussian inhomogeneity and a crescent shape for the sharper inhomogeneity. Resonance with Alfv\'en cyclotron waves diffuses the ion velocities to lower values, giving the crescent distribution its shape. 

 We analyze the effect of an injected turbulent spectrum at one of the boundaries and find it to increase the ion temperature anisotropy only in cases where sub-Alfv\'enic drift was present. In the super-Alfv\'enic case the effect of the turbulence on the ion temperature anisotropy is small compared to the heating due to the drift instability. Nevertheless, the proton temperature anisotropy is decreased by the inclusion of the turbulent spectrum due to increased both parallel and perpendicular energies of ions and protons in the plasma, resulting in higher parallel heating compared to the case without the injected waves, and the corresponding decrease of anisotropy. 

We find that the relaxation of the super-Alfv\'enic drift together with the magnetic turbulence spectrum  injected in the inhomogeneous (Gaussian) background density profile results in magnetic fluctuations fitted by a power-law of the form $f^{-1.5}$. The expansion and the inhomogeneity do not appear to affect the magnetic fluctuation spectrum significantly. An initial super-Alfv\'enic drift alone in the plasma causes magnetic fluctuations with a steeper power law ($-2.2$). 

In conclusion, the  temperature anisotropy for both ions and protons observed in fast wind streams requires perpendicular heating that is stronger than the cooling effects of solar wind expansion. Thus, kinetic instability leading to perpendicular plasma heating must take place in the expanding solar wind close to the Sun and possibly throughout the inner heliosphere, and both, drift instability and ion-cyclotron instability, are good candidates as demonstrated in this study. We analyzed the instability caused by a super-Alfv\'enic drift and the injection of a turbulence spectrum and both cases succeed in producing temperature anisotropies comparable to observations. The inhomogeneity in the plasma enhances the production rate and magnitude of the oblique waves, which in turn contribute to heating. Nevertheless, we find that the parallel propagating left-hand polarized cyclotron waves dominate the resonant heating of the solar wind ions. 

\acknowledgments NO would like to acknowledge support by the Helen Kimmel Center for Planetary Science. LO would like to acknowledge support by NASA grant NNX10AC56G.


\newpage
\newpage

\begin{table}
\begin{tabular}{|c|c|c|c|c|c|c|}
\hline
Case \#  & Exp. param. $\epsilon$ & $V_d$ [$V_A$] & Inhom. param. $q$ & $B_{z0}$ & ($\omega_1,\omega_N$) [$\Omega_p$] & Slope $p$ \\ \hline
1a & 0 & 0.5 & 2 & -- & -- & -- \\
1b & $10^{-4}$ & 0.5 & 2 & -- & -- & -- \\
1c & $10^{-3}$ & 0.5 & 2 & -- & -- & -- \\ \hline
2a & 0 & 0.5 & 6 & -- & -- & -- \\
2b & $10^{-4}$ & 0.5 & 6 & -- & -- & --  \\
2c & $10^{-3}$ & 0.5 & 6 & -- & -- & -- \\ \hline
3a & 0 & 2 & 2 & -- & -- & -- \\
3b & $10^{-4}$ & 2 & 2 & -- & -- & -- \\
3c & $10^{-3}$ & 2 & 2 & -- & -- & -- \\ \hline
4a & 0 & 2 & 6 & -- & -- & -- \\
4b & $10^{-4}$ & 2 & 6 & -- & -- & --  \\
4c & $10^{-3}$ & 2 & 6 & -- & -- & -- \\ \hline
5a & $10^{-4}$ & 2 & 2 & 0.03 & (0.06,0.4) & 1\\ 
5b & $10^{-4}$ & 0 & 2 & 0.03 & (0.06,0.4) & 1\\ 
5c & $10^{-4}$ & 0.5 & 2 & 0.03 & (0.06,0.4) & 1\\ 
5d & $10^{-4}$ & 2 & 6 & 0.03 & (0.06,0.4) & 1\\ 
5e & $10^{-4}$ & 0 & 6 & 0.03 & (0.06,0.4) & 1\\ \hline
6a & $10^{-4}$ & 2 & 2 & 0.03 & (0.06,0.9) & 1\\ 
6b & $10^{-4}$ & 0 & 2 & 0.03 & (0.06,0.9) & 1\\ \hline

\hline
\end{tabular}
\caption{Initial parameters for the different cases described in this study}\label{par_table}
\end{table}
\end{document}